\documentclass[%
 reprint,
 amsmath,amssymb,mathtools,
 aps,
pra,
]{revtex4-2}
\usepackage{booktabs}
\usepackage{graphicx}
\usepackage{subcaption}
\usepackage{dcolumn}
\usepackage{bm}
\usepackage{lipsum}
\usepackage{multirow}
\usepackage{xcolor}
\usepackage{relsize}

\newcommand{\tj}[6]{ \begin{pmatrix}
  #1 & #2 & #3 \\
  #4 & #5 & #6 
\end{pmatrix}}
\begin{document}

\preprint{APS/123-QED}

\title{Rotational quenching of monofluorides in a cryogenic helium bath }

\author{Mateo Londoño }
\author{Jesús Pérez-Ríos}
\affiliation{Department of Physics and Astronomy, Stony Brook University, Stony Brook, New York 11794, USA}


\begin{abstract}
Buffer gas cooling, one of the most relevant direct cooling techniques for cooling molecules, relies on dissipating the energy of the molecule via collisions with a buffer gas. The cooling efficiency hinges on the molecule-atom scattering properties, concretely, on the transport properties. This work presents a global study on the interactions, collision dynamics, and transport properties of monofluoride molecules (X-F), being X a metal, in the presence of a cold He buffer gas. The interactions are calculated using ab initio quantum chemistry methods, and the dynamics is treated fully quantal, assuming the monofluoride molecule is a rigid rotor. The resulting thermalization and rotational quenching rates are analyzed in light of the Born Distorted Wave Approximation (BDWA), yielding an explanation based on the elemental physical properties of the molecule under consideration. Therefore, the analysis of our results reveals the physics behind the rotational quenching of molecules in the presence of a cold buffer gas.


\end{abstract}

--\maketitle

\section{Introduction}
Ultracold polar molecules are a desirable platform to study different aspects of quantum sciences and technologies \cite{DeMille2002, Bohn2017, Kaufman2021, Cornish2024}. The rich landscape of internal degrees of freedom and the presence of long-range dipolar forces make these systems ideal for quantum information processing \cite{Cornish2024} and quantum simulation \cite{carroll2024,schafer2020}, as well as for exploring new pathways to novel quantum materials \cite{Langen2022,Langen2022b}. From a few-body perspective, ultracold molecules have paved the way for studying ultracold chemistry, where explicit quantum effects significantly contribute to reactive processes, offering opportunities for deeper understanding and exploitation \cite{Karman2024, Liu2023, Masato2024}. 

Ultracold molecule samples can be produced via indirect or direct cooling techniques. Indirect cooling refers to tools dedicated to creating cold molecules out of a previously cold ensemble of atoms via photoassociation~\cite{Lett1995,PA,PA2,PAJPR}, magnetoassociation~\cite{Moerdijk,Magnetoassociation} or mergeassociation~\cite{MergeAssociation}. On the contrary, direct cooling techniques start with an ensemble of hot molecules brought to the ultracold regime by dissipating the kinetic energy of the molecules via external fields or through collisions with a reservoir. Among them, we find Stark deceleration~\cite{Stark}, Zeeman slowing~\cite{Zeeman}, centrifuge deceleration~\cite{Cherenkov}, buffer gas cooling~\cite{Weinstein,Buffergas1,Buffergas2} and sympathetic cooling~\cite{sympa1,sympa2,Tobias}, and laser cooling and optoelectrical cooling~\cite{lasercooling,Prehn2016}. Laser cooling is the most prominent technique to bring molecules into the ultracold regime, and it has been proven successful for monofluorides and monohydride molecules \cite{Doyle2016,Zelevinsky2022,Tim2024}. The efficiency of laser cooling is contingent on our ability to have steady and cold sources of molecules, and buffer gas cooling is the best way to achieve it. 

Buffer gas cooling efficiency hinges on the thermalization and energy transfer mechanism of atom-molecule collisions. In this line, there is a wide range of theoretical and experimental studies investigating such aspects in monofluorides, including YbF, AlF, CaF, MgF, and others \cite{Wright2022, Truppe2018, Sangami2024, Karra2022, Skoff2011}. However, there is still no general understanding of the properties that a molecule requires to relax rotationally fast in the presence of a helium buffer gas. Nevertheless, AlF outperforms other monofluorides, even though its properties are similar to any other monofluoride.

In this work, we study the quantum dynamics associated with the rotational quenching of different XF molecules, being X, an alkaline-earth metal or a metal, in a cryogenic helium bath at temperatures ranging from 0.1 K - 10~K. Our results rely on a multi-channel scattering approach, using ab initio potential energy surfaces, to calculate the state-to-state rotational cross sections and transport cross sections of several monofluorides-He collisions. As a result, and after exploring several systems, we derive a general expression capturing the rotational quenching efficiency in terms of general molecular properties as well as the long-range coefficient for the interaction. Therefore, it reveals the fundamentals of the kinetic-rotational energy transfer mechanism in atom-molecule collisions.

The structure of the paper is outlined as follows: Section \ref{sect:PES} discusses the ab initio computation of the potential energy surfaces. Section \ref{sect:scat_theo} details the theory and approximations used to calculate rotational and translational cross sections as well as thermally averaged reaction rates. Section \ref{sect:xs_results} presents the numerical results for rotational inelastic cross sections and reaction rates, interpreting the obtained values using the Born Distorted Wave approximation. Subsequently, Section \ref{sect:transport} covers the results for transport properties. Finally, we summarize the main findings and discuss future prospects in Section \ref{sect:conclusions}.




\section {Potential Energy Surfaces}
\label{sect:PES}

\begin{figure}[h]
\centering
\includegraphics[scale=0.25]{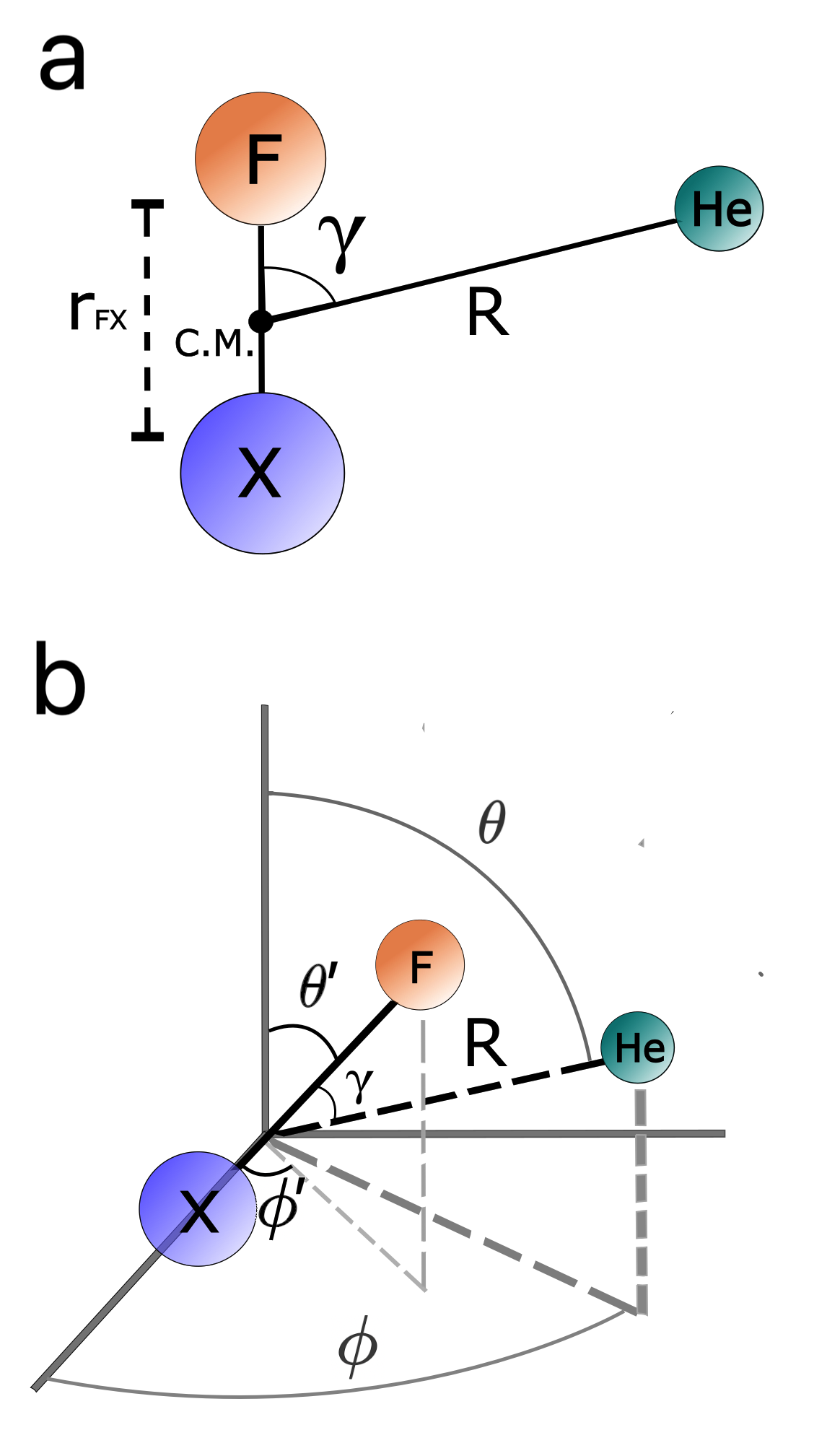}
\caption{Atom-Molecule system. (a) relative orientation of the XF-He system with the Jacobi coordinates (R,$\gamma$) used to compute the interaction potential. (b) Coordinate system used in the rotational dynamics}
\label{fig:atom-mol-system}
\end{figure}

For the simulation of rotational inelastic processes, we assume that the vibrational degrees of freedom do not play a role. As a consequence, the interaction potentials of XF-He or NaH-He depend on two degrees of freedom, as shown in panel (a) of Fig.\ref{fig:atom-mol-system}, one associated to the distance of the vector joining the center of mass of the diatomic molecule and He atom $R$, and the other is the relative angle of such vector with respect to the diatomic axis $\gamma$. 

For CaF($^2\Sigma$)-He and Sr(X$^2\Sigma$)F-He, we have calculated the potential energy surfaces (PESs) using ab initio quantum chemistry methods. We calculate the interaction energy at every given geometry using the coupled cluster with singlet, doubles and perturbative triple excitations approach as implemented in MOLPRO package. For CaF(X$^2\Sigma$)-He the distance between Ca and F atoms is fixed to 1.967~\AA~\cite{database}, whereas for SrF(X$^2\Sigma$)-He the distance between Sr and F atoms is fixed to 1.967~\AA~\cite{database}. The basis set of choice is the AVQZ of Dunning~\cite{basisset}, reaching a good compromise between accuracy and computational cost. Furthermore, the interaction energies are corrected from the basis set superposition error. The PESs consist of 1460 geometries involving 73 values of $R$ between 2.5 and 30~\AA, and 20 angles, 18 of which are associated with the Gauss-Legendre quadrature points and to $\theta=0$ and 180$^{\circ}$.

\begin{figure*}[t] %
    \centering    \includegraphics[width=\textwidth]{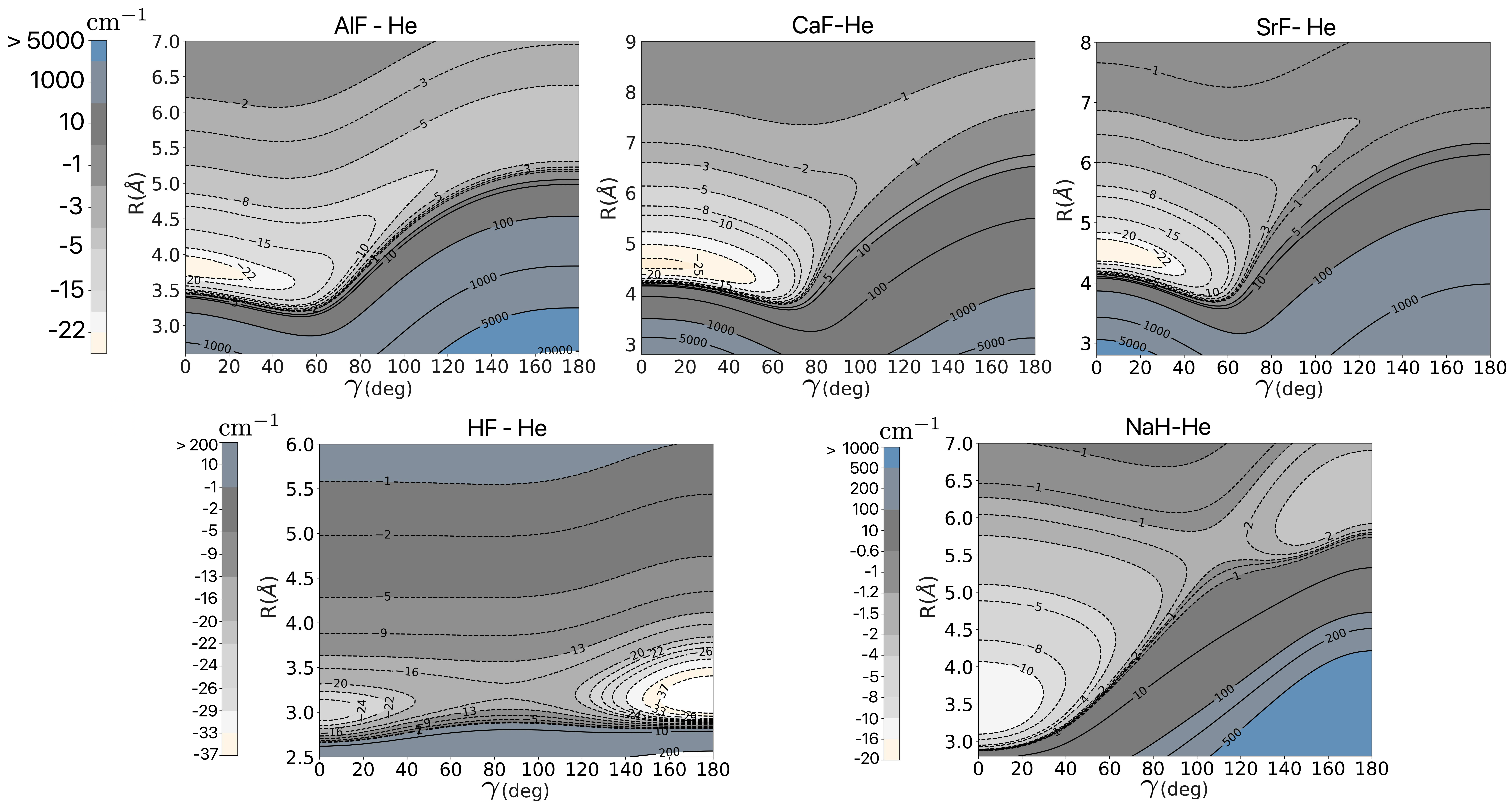} 
    \caption{Potential energy surfaces for the different atom-diatom systems in the Jacobi coordinates depicted in Fig.~\ref{fig:atom-mol-system}. The energy is given cm$^{-1}$.}
    \label{fig:PES}
\end{figure*}

Figure~\ref{fig:PES} shows the PESs for the CaF(X$^2\Sigma$)-He and SrF(X$^2\Sigma$)-He systems. The PES for CaF(X$^2\Sigma$)-He shows a global minimum of of -25.08 cm$^{-1}$ at $R$ = 4.63 $\AA$ and $\gamma = 0$, leading to the only stable linear configuration CaF-He. The same behavior is noticed for SrF(X$^2\Sigma$)-He, although in this case, the minimum is -23 cm$^{-1}$ at R = 4.52 $\AA$. 

In the same Figure, for comparison, we show the PESs for AlF(X$1\Sigma$)-He, HF(X$^1\Sigma$)-He and NaH(X$^1\Sigma$)-He, taken from references \cite{Karra2022}, \cite{Alpizar2024}, \cite{Bop2019} respectively; since we will study them too along this work. As a result, we notice that the energy landscape of every X-F system where X is a metal or an alkaline-earth metal is very similar, just showing small variations. Nevertheless, the global minimum position shifts toward larger distances as the mass of the X atom increases. However, for HF(X$^1\Sigma$)-He, the PES shows two minima, meaning that the He atom can attach to the H or F end of the diatomic molecules. Finally, and surprisingly enough, the PES of NaH(X$^1\Sigma$)-Looks very similar to the X-F systems studied. As shown below, the PES discrepancies between the different systems result in very distinct quenching properties and state-to-state cross sections.

\section{Quantal Scattering Calculations}
\label{sect:scat_theo}
\begin{figure}[h]
\centering
\includegraphics[scale=0.35
]{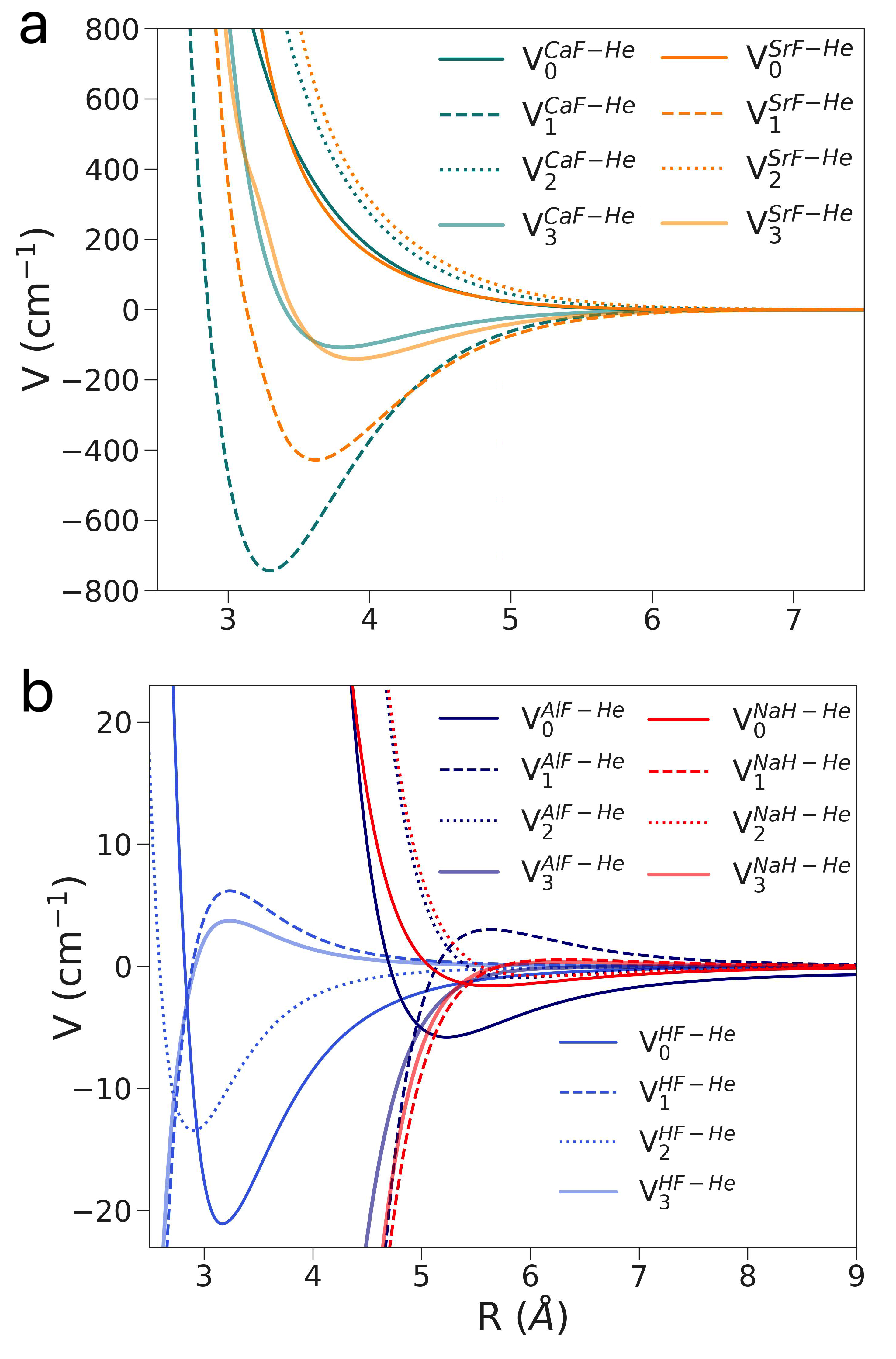}
\caption{The first four radial coefficients V$_{\lambda}$(R) of the CaF($X^{1}\Sigma^{+}$)-He (S$^{1}$) interaction potential energy surface}
\label{fig:rad_coeff-CaF}
\end{figure}

For quantal scattering calculations it is convenient to expand the PES into Legendre polynomials as
\begin{equation}
    V(R,\gamma) = \sum_{\lambda = 0}^{\lambda_{max}} V_{\lambda}(R)P_{\lambda}(\cos \gamma),
    \label{eq:leg_expansion}
\end{equation}
where $P_{\lambda}(x)$ stands for the Legendre polynomial of degree $\lambda$ and argument $x$, and the expansion coefficients are given by 
\begin{equation}
\label{eq:eq2}
V_{\lambda}(R) = \bigg( \frac{2\lambda +1}{2} \bigg)\int _{0}^{\pi}V(R,\gamma)P_{\lambda}(\cos \gamma)\sin \gamma d\gamma.
\end{equation}
The first three radial coefficients for the PESs XF-He, HF-He, and NaH-He, computed through equation \ref{eq:eq2} are shown in Figure~\ref{fig:rad_coeff-CaF}. The main contribution comes from the odd terms. The even terms are not completely dissociative, but the depth is negligible compared to the odd terms. In particular, $V_{\lambda = 1} (R)$ is the most important contribution. However, it is worth noticing that even though the PESs of AlF-He and, CaF-He and SrF-He looked alike, the $V_{\lambda}(R)$ are quite different. For instance, for the isotropic term, $V_{0}(R)$, the minimum appears around 5~\AA~for AlF-He, whereas for CaF-He and SrF-He is around 3.8~\AA. In the same vein, the depth of the well of that term of the radial potential is much shallower in AlF-He than in the other two cases. 

\subsection{Atom-molecule scattering theory}
The evaluation of the atom-molecule scattering event is carried out using the rigid rotor-atom scattering approach developed by Arthurs and Dalgarno \cite{Arthurs1960}, while neglecting the hyperfine structure.

In the center-of-mass frame, considering atomic units, the dynamics of the atom interacting with a rigid-rotor-diatom is described by the Hamiltonian
\begin{equation}
    \hat{H} = -\frac{\nabla^{2}_{R}}{2\mu} + \frac{\hat{L}^{2}}{2\mu R^{2}} + V(R,\gamma) + \hat{H}_{\rm{rot}}.
\end{equation}
The first term refers to the relative kinetic energy component along the scattering coordinate R, with the atom-diatom reduced mass $\mu$. The second term is the centrifugal component of the relative kinetic energy, with angular momentum $l$. The interaction potential $V(R,\gamma)$ are given by the atom-diatom PES in the Jacobi coordinates (see figure \ref{fig:atom-mol-system}). Finally, 
\begin{equation*}
    \hat{H}_{\rm{rot}}Y_{j}^{m}(\theta^{\prime}, \phi^{\prime}) = [B_{0}j(j+1) - D_{e}(j(j+1))^{2}]Y_{j}^{m}(\theta^{\prime}, \phi^{\prime}) ,
\end{equation*}
describes the rigid-rotor dynamics, where $Y_{j}^{m}(\theta^{\prime}, \phi^{\prime})$ represent the spherical harmonics with rotational quantum number $j$ and projection in the quantization axis $m$, $B_{0}$ is the rotational constant of the vibrational ground state of the molecule and $D_e$ is the centrifugal distortion constant at equilibrium.

The anisotropy of interaction potential $V(R,\gamma)$ is the responsible of the rotational transitions. The interaction potential couple scattering channels with rotational quantum numbers $j$ , $j^{\prime}$ and partial waves $l$ and $l^{\prime}$. In the absence of external fields, since $J$ and $M$ are good quantum numbers (conserve quantities)  only those channels ($l,j$) and ($l^{\prime},j^{\prime}$) preserving those two quantum numbers are coupled.

The scattering wave-function $u_{j^{\prime},l^{\prime}}^{Jjl}(R)$ satisfies the Schrödinger equation
\begin{equation}
\label{eq4}
\begin{split}
    \frac{1}{2\mu}&\bigg[\frac{d^{2}}{dR^{2}} - \frac{l^{\prime}(l^{\prime} + 1)}{R^{2}} + k_{jj^{\prime}}^{2}\bigg]u^{Jjl}_{j^{\prime}l^{\prime}}(R)\\
    & = \sum_{j^{\prime \prime} l^{\prime \prime}}V_{j^{\prime \prime} l^{\prime \prime};j^{\prime } l^{\prime } }^{J}(R)u_{j^{\prime \prime} l^{\prime \prime}}^{J j^{\prime } l^{\prime }}(R),
\end{split}
\end{equation}
where the channel wave number
\begin{equation}
\begin{split}
        k^{2}_{jj^{\prime}} &= 2\mu ( E_{k} +B_{0}j(j+1) - D_{e}[j(j+1)]^{2} \\
        & -B_{0}j^{\prime}(j^{\prime}+1) + D_{e}[j^{\prime}(j^{\prime}+1)]^{2}
\end{split} 
\end{equation}
depends on the relative kinetic energy $E_{k}$, or collision energy. In Eq.(\ref{eq4}), by means of Eq.(\ref{eq:leg_expansion}), we find
\begin{equation}
\begin{split}
        V^{J}_{j'l';jl}(R) &= \sum_{\lambda}^{\lambda_{\text{max}}} V_{\lambda}(R)(-1)^{J+l'+l} \\
&\sqrt{(2j+1)(2l+1)(2j'+1)(2l'+1)(2\lambda+1)}\\
&
\begin{pmatrix}
l & \lambda & l' \\
0 & 0 & 0
\end{pmatrix}
\begin{pmatrix}
j & \lambda & j' \\
0 & 0 & 0
\end{pmatrix}
\left\{
\begin{matrix}
j & \lambda & j' \\
l & J & l'
\end{matrix}
\right\},
\end{split}
\label{eq:pot_mat_elem}
\end{equation}
with (.) and \{{.}\} representing the 3j and 6j-symbols respectively. Note that the symbols impose the selection rule $\Delta l = \lambda$ or $\Delta j = \lambda$.

The set of coupled-channels equations in (\ref{eq4}) are solve numerically, and the solution matched with the analytical asymptotic expression for the radial wave-function 
\begin{equation*}
\begin{split}  
    u^{JM}_{jl,j'l'}(R \to \infty) \sim & \delta_{jj'}\delta_{ll'} \sin\left(k_j R - l' \frac{\pi}{2}\right)\\
& + \frac{e^{i(k_j R - l' \pi / 2)}}{\sqrt{k_{j'}}}T^{JM}_{jl,j'l'},
\end{split}
\end{equation*}
getting the transition matrix $T^{JM}_{jl,j'l'}$, which accounts for the probability of going from channels ($l,j$) to ($l^{\prime},j^{\prime}$) during the collision. This matrix allow us to compute the state-to-state cross sections
\begin{equation}
    \sigma_{j \to j'}(E_k) = \frac{4\pi}{k^2_{jj} (2j+1)} \sum_{Jl l'} (2J+1) \left|T^{JM}_{j;l;j'l'}\right|^2,
\end{equation}
and, with it, reaction rates. 


\subsection{Transport cross-section}
Equally important to the inelastic cross section discussed earlier, the transport cross section offers critical insights into the translation of cooled target molecules within a buffer gas cell, facilitating the generation of a steady flow of cold molecules. In particular, we pay attention to the diffusive and viscosity cross sections, to elucidate the efficiency of the transport of cold molecules during the thermalization in buffer gas. 

Quantum mechanically, these two transport cross sections can be computed as \cite{mottbook}
\begin{equation}
    \sigma_{D}(E_{k}) = \frac{4\pi}{k^{2}} \sum_{l = 0}^{\infty} (l + 1) \sin^{2}(\delta_{l+1}(E_{k}) - \delta_{l}(E_{k}))
    \label{eq:dxs_sc}
\end{equation}
and,
\begin{equation}
    \sigma_{\eta}(E_{k}) = \frac{4\pi}{k^{2}}\sum_{l = 0}^{\infty} \frac{(l+1)(l+2)}{2l+3} \sin^{2}(\delta_{l+2}(E_{k}) - \delta_{l}(E_{k})).
    \label{eq:vxs_sc}
\end{equation}
$\delta_{l}(E_{k})$ is the scattering phase shift for a given partial wave $l$ and collision energy $E_{k}$ in the single channel model and $k^{2} = 2\mu E_{k} $. For a single scattering channel the elastic scattering cross section is
\begin{equation}
    \sigma_{el} (E_{k}) = \frac{4\pi}{k^{2}}\sum_{l=0}^{\infty}(2l+1)\sin^{2} (\delta_{l}(E_{k})).
    \label{eq:exs_sc}
\end{equation}

Transport cross sections can be averaged over the thermal ensemble in order to observe the net propagation of molecular beams in the experiments, we compute such averages by evaluating the integrals 

\begin{equation}
    \sigma_D(T) = \frac{1}{2(k_B T)^3} \int_0^\infty \sigma_D(E_k) \exp\left[-\frac{E_k}{k_B T}\right] E_k^2 \, dE_k,
\end{equation}
and,
\begin{equation}
    \sigma_\eta(T) = \frac{1}{6(k_B T)^4} \int_0^\infty \sigma_\eta(E_k) \exp\left[-\frac{E_k}{k_B T}\right] E_k^3 \, dE_k,
\end{equation}

\begin{table*}[t]
\centering
\caption{Parameters used for the coupled-channel quantum scattering calculations. DR here stands for the spatial step size, OTOL and DTOL refer to the off-diagonal and diagonal tolerance thresholds for the cross-section convergence, and NLEVEL is the number of terms used in the potential expansion.}
\begin{tabular*}{\textwidth}{@{\extracolsep{\fill}}lccccc}
\toprule
\hline
\textbf{Molecule} & \textbf{AlF} & \textbf{CaF} & \textbf{SrF} & \textbf{HF} & \textbf{NaH} \\
\midrule
$\mu$ (a.m.u.) & 3.68207364173 & 3.74861835234 & 3.1862378 &  3.36192483 & 3.430440591 \\
$B_0$ (cm$^{-1}$) &0.5499923150168 & 0.3385 &0.25053 & 20.9557 & 4.89 \\
$D_e$ ($\times$10$^{-7}$cm$^{-1}$) &10.40719977 & 0.45 & 2.49 & 21.51 &  \\
\\
\hline
\hline
\\
$[R_{\text{min}}, R_{\text{max}}]$ (\AA) & & \hspace{20 px}[2.5, 200] & & & \\
DR (\AA) & &\hspace{20 px} 0.006 & & & \\
OTOL (\AA$^2$) & & \hspace{20 px}0.001 & & & \\
DTOL (\AA$^2$) & &\hspace{20 px} 0.1 & & & \\
NLEVEL & &\hspace{20 px} 12 & & & \\
\bottomrule
\label{tab:MOLSCAT_param}
\end{tabular*}
\end{table*}
 
\subsection{Distorted-wave treatment}
\label{DWBA}
The Born Distorted-Wave Approximation (BDWA) provides a framework for calculating inelastic scattering cross sections when a molecule undergoes a rotational state change due to a collision. Within the BDWA , the potential is split in two contributions as~\cite{Arthurs1960,messiah61}
\begin{equation}
    V(R, \gamma) = U(R, \gamma) + W(R, \gamma),
\end{equation}
where $U(R,\gamma)$ is the distorted potential in the rotational state $j$ and $W(R, \gamma)$ the perturbation generating the rotational transition $j\rightarrow j^{\prime}$. Considering the expansion of the total potential as shown in Eq.~(\ref{eq:leg_expansion}), the distorted and perturbative potentials can be written as
\begin{equation}
    U(R,\gamma) = V_{j}(R)P_{j} (\cos \gamma) \,\,\, ; \,\,\, W(R,\gamma) = V_{\lambda}(R) P_{\lambda} (\cos \gamma).
\end{equation}
and after taking into account the selection rules coming from the matrix elements given by Eq.~(\ref{eq:pot_mat_elem}), we find $j= j^{\prime}  + \lambda$. A zeroth-order solution comes from neglecting the couplings in the right-hand side of equation (\ref{eq4}) which yields to the homogeneous single channel equation
\begin{equation}
    \bigg[\frac{1}{2\mu}\bigg(\frac{d^{2}}{dR^{2}} - \frac{l^{\prime}(l^{\prime} + 1)}{R^{2}} + k_{jj^{\prime}}^{2}\bigg) + V_{ j l;j^{\prime } l^{\prime } }^{J}(R)\bigg]u^{Jjl}_{j^{\prime}l^{\prime}}(R)=0,
    \label{eq:DWA0}
\end{equation}
with solution 
\begin{equation}
    u_{j^{\prime}l^{\prime}}^{jlJ} = \delta_{j^{\prime}j}\delta_{l^{\prime}l}w_{j^{\prime}l^{\prime}}^{jlJ}
\end{equation}
which asymptotically behaves as
\begin{equation}
w_{j^{\prime}l^{\prime}}^{jlJ} \sim \sin(k_{jj^{\prime}}r - \frac{l^{\prime}\pi}{2} + \eta_{j^{\prime}l^{\prime}}^{jlJ}).
\end{equation}
The first order correction is obtained when solving the equation 

\begin{equation}
\begin{split}
    \bigg[\frac{\hbar^{2}}{2\mu}\bigg(\frac{d^{2}}{dR^{2}} - &\frac{l^{\prime}(l^{\prime} + 1)}{R^{2}} + k_{jj^{\prime}}^{2}\bigg) + V_{j^{ \prime} l^{\prime};j^{\prime } l^{\prime } }^{J}(R)\bigg]u^{Jjl}_{j^{\prime}l^{\prime}}(R) \\
    & = W_{jl;j^{\prime } l^{\prime }}^{J}(R)(\delta_{j^{\prime}j}\delta_{l^{\prime}l} - 1)w_{jl}^{J}(R).
    \label{eq:DWA1}
    \end{split}
\end{equation}
Then, the S-matrix elements are given by 
\begin{equation}
    S^{J}_{jl;j^{\prime } l^{\prime } } = e^{i(\eta_{j^{\prime}l^{\prime}}^{jlJ} + \eta_{jl}^{jlJ})}\Bigg( \frac{k_{j^{\prime}j}}{k_{j^{\prime}j}}\Bigg)^{1/2}\bigg[ \delta_{jj^{\prime}}\delta_{ll^{\prime}} \big(1-2\rm{i}\beta_{j^{\prime}l^{\prime}}^{jlJ}\big) + 2\rm{i}\beta_{j^{\prime}l^{\prime}}^{jlJ}\bigg]
\end{equation}
with the coupling coefficients
\begin{equation}
\label{eq20}
    \beta_{j^{\prime}l^{\prime}}^{jlJ} = \int_{0}^{\infty} w_{j^{\prime}l^{\prime}}^{jlJ}(R)W_{ jl;j^{\prime } l^{\prime }}^{J}(R)w_{jl}^{jlJ}(R)dR.
\end{equation}

In order to obtain the state-to-state cross section, $\sigma_{j\rightarrow j^{\prime}}$, it is mandatory to evaluate Eq.~(\ref{eq20}), which involved the matrix elements
\begin{equation}
    \begin{split}
       & W_{\lambda \,\, jl;j^{\prime } l^{\prime }}^{J}(R)  = 4\pi  V_{\lambda}(R)\mathlarger{\sum}_{\substack{m_{\lambda},m_{j,j^{\prime}}\\
       m_{l,l^{\prime}}}}\frac{C^{Jjl*}_{m_{j}m_{l}M} C^{Jj^{\prime}l^{\prime}}_{m_{j}^{\prime}m_{l}^{\prime}M}}{\sqrt{2\lambda + 1}} \\
                  &\times \tj{\lambda}{l}{l^{\prime}}{0}{0}{0}\tj{\lambda}{l}{l^{\prime}}{m_{\lambda}}{m_{l}}{m_{l^{\prime}}}\tj{\lambda}{j}{j^{\prime}}{0}{0}{0}\tj{\lambda}{j}{j^{\prime}}{m_{\lambda}}{m_{j}}{m_{j^{\prime}}}\\
                  & \times A_{\lambda l l^{\prime}}A_{\lambda j j^{\prime}}
                  \end{split}
\end{equation}
where $C^{Jjl*}_{m_{j}m_{l}M}$ are the Clebsh-Gordan coefficients and $A_{l_{1} l_{2} l_{3}}=\sqrt{(2l_{1} +1)(2l_{2} +1)(2l_{3} +1)/4\pi}$.
This last equation can be used to evaluate the S-matrix elements and the state-to-state cross section, yielding
\begin{equation}
    \begin{split}
        &\sigma_{j\rightarrow j^{\prime}}^{\rm{BDWA}}(E_k)= \frac{4\pi}{(2j+1)k_{jj}^{2}}\frac{k_{j^{\prime}j}}{k_{jj}}\mathlarger{\sum_{J=0}^{\infty}}(2J+1)\mathlarger{\sum_{l = |J-j|}^{J+j}}\mathlarger{\sum_{l^{\prime} = |J-j^{\prime}|}^{J+j^{\prime}}}\\
        &\Bigg( 4\pi  \mathlarger{\sum}_{\substack{m_{\lambda},m_{j,j^{\prime}}\\
       m_{l,l^{\prime}}}}\frac{C^{Jjl*}_{m_{j}m_{l}M} C^{Jj^{\prime}l^{\prime}}_{m_{j}^{\prime}m_{l}^{\prime}M}}{\sqrt{2\lambda + 1}} \tj{\lambda}{l}{l^{\prime}}{0}{0}{0}\tj{\lambda}{l}{l^{\prime}}{m_{\lambda}}{m_{l}}{m_{l^{\prime}}} \\
       & \times\tj{\lambda}{j}{j^{\prime}}{0}{0}{0}\tj{\lambda}{j}{j^{\prime}}{m_{\lambda}}{m_{j}}{m_{j^{\prime}}}
                   \times A_{\lambda l l^{\prime}}A_{\lambda j j^{\prime}}\Bigg)^{2}\big(\mathcal{R}_{j^{\prime}l^{\prime}}^{Jjl}\big)^{\,2},
    \end{split}
    \label{eq:BDWA_xs}
\end{equation}
with
\begin{equation}
    \mathcal{R}_{j^{\prime}l^{\prime}}^{Jjl} = \int_{0}^{\infty} w_{j^{\prime}l^{\prime}}^{jlJ}(R)V_{\lambda}(R)w_{jl}^{jlJ}(R)dR.
    \label{eq:BDWA_rad}
\end{equation}

\section{Computational details}

In order to compute the rotationally inelastic cross section, we use the package MOLSCAT \cite{HUTSON2019} under the rigid-rotor approximation of Dalgarno $et. al$ (see sec \ (3)). The coupled-channel equations were solved using the log-derivative method of Manolopolous \cite{Manolopoulos1986} combined with the Airy propagator for long range \cite{Alexander1984}, between 1.8$\AA$ and 200$\AA$, and a time step of 0.008 $\AA$. The number of partial waves for the calculations was converged automatically by the code setting the option JTOTU = 99999, and the total number of rotational states included were converged for every atom+molecule case, but in average, a total of 12 states were included. It means j = 0 - 11. A summary of the simulation parameters for all the systems is shown in table \ref{tab:MOLSCAT_param}.

The transport cross sections  (equations \ref{eq:dxs_sc} and \ref{eq:vxs_sc}) and the elastic single-channel quantal cross section (equation \ref{eq:exs_sc}) were obtained by solving the Schrödinger equation using Numerov's method. Calculations were performed with a grid of 10$^{5}$ steps, spanning from R$_{min}$ = 2.5 $a_{0}$ to R$_{max}$, where R$_{max}$ = 400$a_{0}$ for low collision energies and R$_{max}$ = 80$a_{0}$ for high collision energies. The number of partial waves was selected to ensure a convergence of 1\% in the cross sections, requiring $\le$ 20 partial waves. Only the isotropic component of the potential energy surface, $V_{0}(R)$, was considered for these calculations.

\section{Results}
\label{sect:xs_results}

\subsection{State-to-state cross sections}

\begin{figure}[h]
\centering
\includegraphics[scale=0.25]{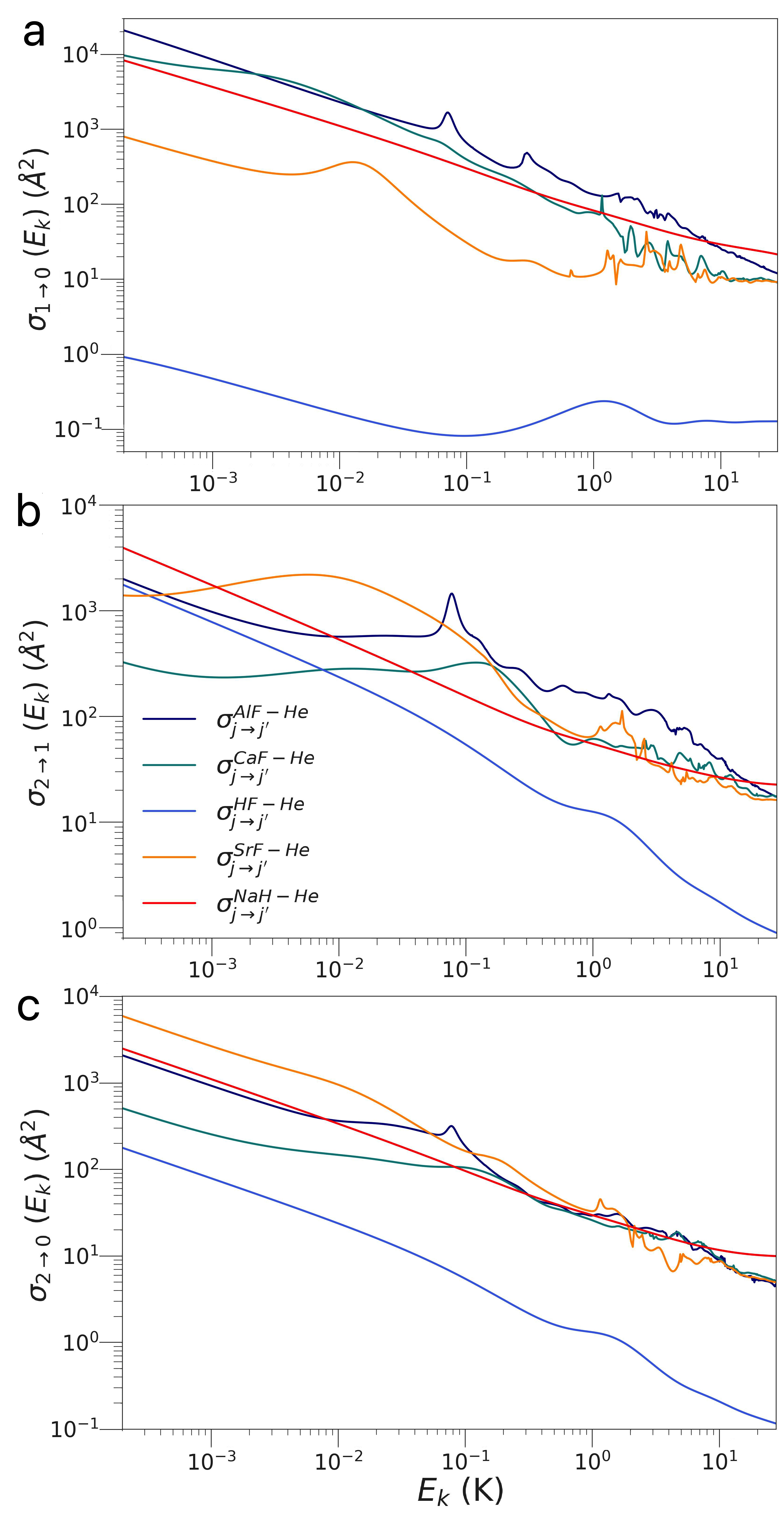}
\caption{State-to-state rotational cross section for all the systems explored in this work as a function of the collision energy. Every color is associated with a given atom-molecule system.}
\label{fig:ixs_j012}
\end{figure}

Figure~\ref{fig:ixs_j012} displays the state-to-state rotational cross section for processes: XF(j)+He$\rightarrow$ XF(j')+He, with $j'<j$, with X=H, Al, Ca, Sr and for NaH(j)+He$\rightarrow$ NaH(j')+He. In general, by virtue of the Wigner threshold law, we notice that the inelastic cross-section is inversely proportional to the collision energy at low collision energies. In the case of AlF-He, CaF-He, and SrF-He, that trend is perturbed by broad shape resonances below 100~mK, followed by the thermal region, at which many partial waves contribute to the scattering, yielding lots of narrow shape resonances. At this region, the state-to-state cross sections are very similar due to the resemblance of PESs of these systems, as shown in Figure~\ref{fig:PES}. On the contrary, the state-to-state cross section for NaH+He and HF+He show a relatively smooth behavior compared to the rest of the systems; almost no resonances are present. This is expected since NaH and HF are lighter than the other XF molecules; hence, they will show fewer bound states and resonances for similar interaction potentials. The situation is even more critical for NaH-He since the PES shows a shallow minimum of around 10 cm$^{-1}$.

$\sigma_{1\rightarrow 0}(E_k)$ is the most relevant state-to-state cross section to establish the rotational quenching efficiency of a molecule in a buffer gas. In this case, as displayed in panel (a) of Figure~\ref{fig:ixs_j012}, we notice that AlF-He shows the largest cross section, whereas HF shows the smallest. It makes sense that HF shows the smallest $\sigma_{1\rightarrow 0}(E_k)$ value since it shows the most isotropic PES, as shown in  Figure~\ref{fig:PES}. However, it is unclear why AlF-He shows a larger cross section than the rest of the systems, even though the PES look alike. To elucidate this further, we will look into the reaction rates instead.

\subsection{State-to-state rate constants}

\begin{figure}[h]
\centering
\includegraphics[scale=0.25
]{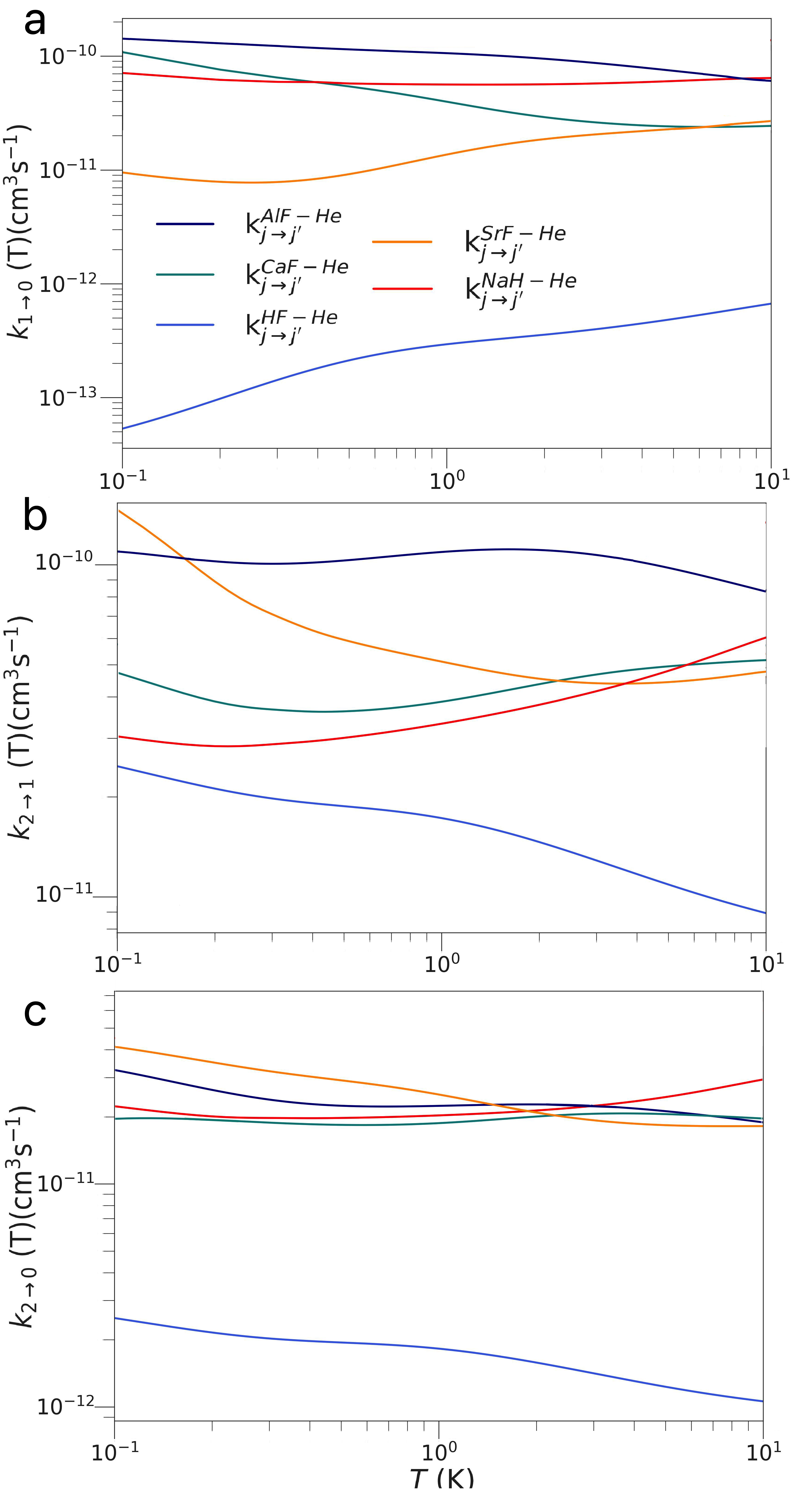}
\caption{Temperature dependent state-to-state reaction rates for all the systems explored in this work. The color code is the same as in Fig.~\ref{fig:ixs_j012}.}
\label{fig:rr_012_all}
\end{figure}

Next, we compute the state-to-state rotational rate constants $k_{j\rightarrow j^{\prime}}$ as
\begin{equation}
    k_{j\rightarrow j^{\prime}} (T) = \Bigg(\frac{8}{\pi\mu \beta^{3}} \Bigg)^{1/2}\int_{0}^{\infty}\sigma_{j\rightarrow j^{\prime}}(E_{k})\exp(-E_{k}/\beta)E_{k}dE_{k},  
\end{equation}
where $\beta = k_{B}T$, being $k_B$ the Boltzmann constant. The results for the state-to-state reaction rate are displayed in Figure \ref{fig:rr_012_all} for temperatures between 0.1 and 10~K, for the same rotational transitions as in Figure~\ref{fig:ixs_j012}. Note that the resonances are washed out due to the thermal averaging, and the rates show a flat behavior. All the systems under consideration yield the same order of magnitude for the rate constant for a given inelastic process, except HF-He, due to the isotropic nature of the PES and the large rotational constant of HF. Nevertheless, for $k_{1\rightarrow 0}(T)$ and $k_{2\rightarrow 1}(T)$, AlF+He shows a larger rate constant than the rest of XF-He systems. On the contrary, for $k_{2\rightarrow 0}(T)$, the rate constants are almost identical. Therefore, the rates are almost identical in processes involving the exchange of two-rotational quanta, $k_{2\rightarrow 0}(T)$. In other words, the $V_{2}(R)$ term of the PES is very similar in all XF systems, as can be seen in Fig.~\ref{fig:rad_coeff-CaF}.

The rotational quenching efficiency of a molecule in a buffer gas is mainly controlled by  $k_{1\rightarrow 0}(T)$, since it is the most relevant state-to-state rate for the rotational relaxation time~\cite{Montero}. Table \ref{tab:k10} shows the values for the $k_{1\rightarrow 0}(T)$ for a few temperatures relevant to buffer gas cooling experiments, indicating that, indeed, AlF rotationally quenches more efficiently than the rest of XF molecules even though the underlying PES for all the XF-He systems is very similar. On the contrary, the $V_{1}(R)$ interaction term is very different in AlF-He from SrF-He and CaF-He, explaining the large rotational quenching efficiency of AlF.

\begin{table}
\begin{tabular}{l c c c} 
\hline\hline   

 system & T = 0.5 K & \hspace{4 pt} T = 4 K & \hspace{4
 pt} T = 10 K  
\\ [0.5ex]  
\hline   
\\
AlF+He & \hspace{2pt} 1.29 $\times$ 10$^{-10}$& \hspace{10 pt}  9.09 $\times$ 10$^{-11}$& \hspace{14pt} 7.1 $\times$ 10$^{-11}$\\[1 ex]
CaF+He & \hspace{2pt} 6.33$\times$ 10$^{-11}$& \hspace{10 pt}  2.90$\times$ 10$^{-11}$& \hspace{14pt} 2.90 $\times$ 10$^{-11}$\\[1 ex]
SrF+He & \hspace{2pt} 1.28 $\times$ 10$^{-11}$& \hspace{10 pt}  2.83 $\times$ 10$^{-11}$& \hspace{14pt} 2.69 $\times$ 10$^{-11}$\\[1 ex]
HF+He & \hspace{2pt} 3.68 $\times$ 10$^{-13}$& \hspace{10 pt}  4.89 $\times$ 10$^{-13}$& \hspace{14pt} 6.68 $\times$ 10$^{-13}$\\[1 ex]
NaH+He & 5.96 $\times$ 10$^{-11}$ &\hspace{10pt}6.09 $\times$ 10$^{-11}$& \hspace{16pt}6.58 $\times$ 10$^{-11}$ \\[1 ex]
\hline 
\end{tabular} 
\caption{Reaction rate coefficient $k_{1\rightarrow0}$ in cm$^{3}$s$^{-1}$ for different AB-He systems at different temperatures}
\label{tab:k10}
\end{table}  

\subsection{Universal trends on rotational quenching}

Rotational quenching in atom-molecule collisions depends on the coupling strength between the translational degrees of freedom and the rotational ones of the molecule. This is intimately related to the anisotropy of the underlying PES. In the case of a molecule in a buffer gas, due to the cryogenic environment, the figure of merit is the relaxation rate from $j=1$ to $j=0$, given by $k_{1\rightarrow 0}(T)$. A larger relaxation rate indicates an efficient energy transfer mechanism between the translational energy and the molecule's internal rotational energy. In that respect, having the state-to-state rate constants for five different molecule-He systems (see Fig.~\ref{fig:rr_012_all}) brings us the opportunity to elucidate the physics behind rotational quenching and to understand why AlF-He is different from the rest of XF-He systems. To this end, we use the BDWA framework presented in Section~\ref{DWBA} considering $\lambda\rightarrow 0$ transitions, consistent with the exchange of $\lambda$ rotational quanta. Then, assuming the long-range tail of the molecule-atom PES as 

\begin{equation}
V_{\lambda}(R) =  \frac{C_{6}^{(\lambda)}}{R^{6}},
\end{equation}
one finds that the reaction rate $k_{\lambda\rightarrow 0 }$ is proportional to  
\begin{equation}
    k_{\lambda \rightarrow 0} \propto  \sqrt{\frac{B_{0}}{\mu \beta^{3}}}(C^{(\lambda)}_{6})^{2}.
    \label{eq:scaling_law}
\end{equation}

\begin{figure}[htbp]
\centering
\includegraphics[scale=0.38
]{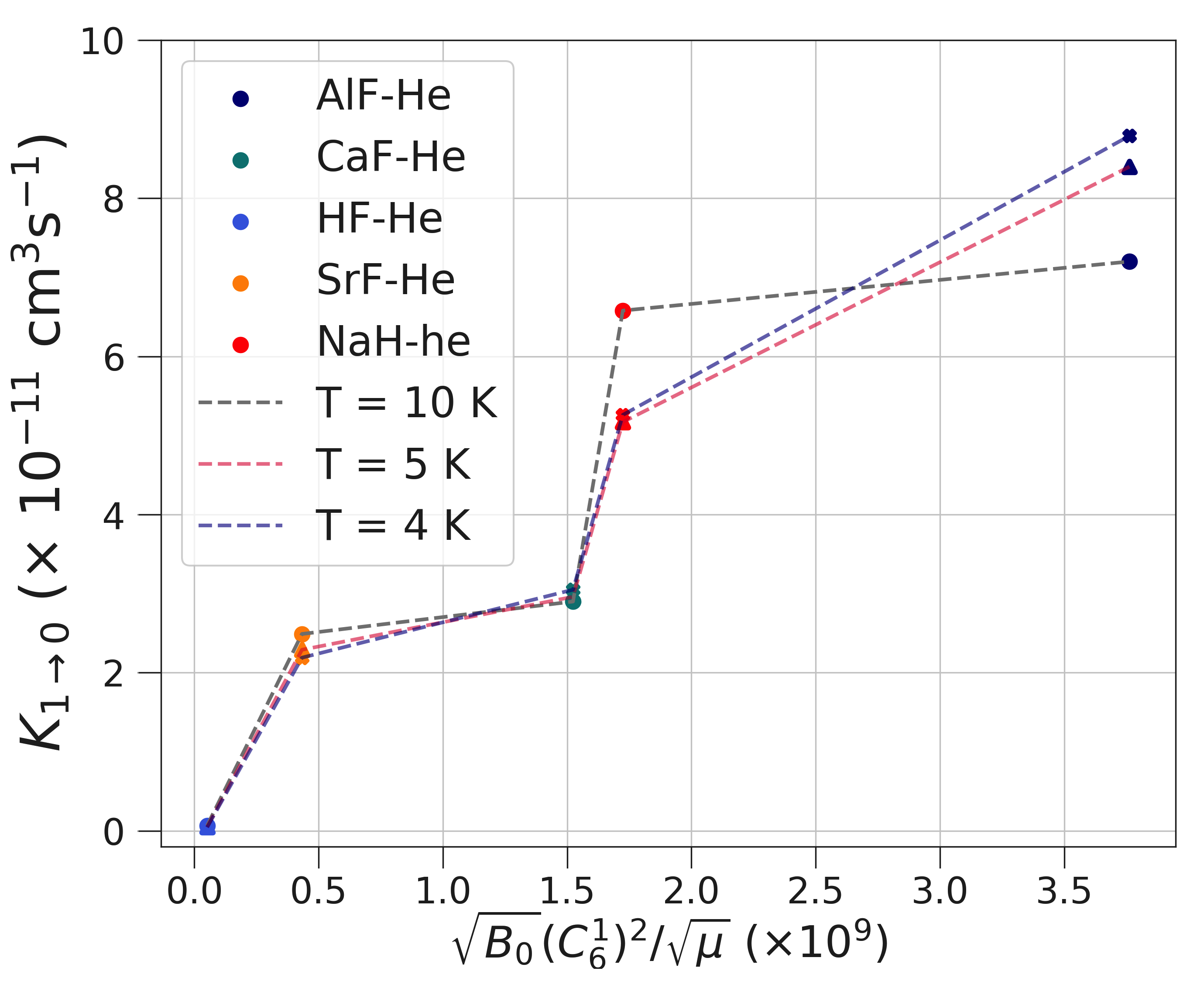}
\caption{State-to-state rate constants $k_{1\rightarrow 0}(T)$ versus the prescribed scaling law in Eq.~(\ref{eq:scaling_law}).}
\label{fig:scaling_law}
\end{figure}

To test the validity of Eq.~(\ref{eq:scaling_law}), we compare the scattering results for 
$k_{1\rightarrow 0}(T)$ versus Eq.~(\ref{eq:scaling_law}), and the results are displayed in Figure \ref{fig:scaling_law}. Our scaling requires the $C_{6}^{1}$ coefficients, tabulated in Table~\ref{tab:table_BDWA}, and derived from the computed long-range tail of the calculated PES, along as the reduced mass of the molecule-He system and rotational constant of the molecule in the vibrational ground state, both parameters shown in the Table. The scaling law predictions are tested against the scattering data, and the results are displayed in Figure \ref{fig:scaling_law} for the transition $\lambda=1 \rightarrow0$ at different temperatures. The figure shows that our scaling law properly captures the hierarchy for the different systems, even at different temperatures.

From Table~\ref{tab:table_BDWA}, one notices that AlF, SrF and CaF have very similar values for $ B_{0} $ and $ \mu $, but $ C_{6}^{1} $ is significantly larger for AlF than for CaF or SrF, yielding a larger relaxation rate. This suggests that differences in the first anisotropic term of the potential, rather than the similarities in rotational constant and reduced mass, play the most significant role. In that vein, from an electronic structure standpoint, Ca and Sr belong to Group~2 (alkaline earth metals) with fully occupied $ s $-states, whereas Al, from Group~13, has an unpaired electron in its $ 3p $-state, which could explain the prominent $V_{1}(R)$ interaction potential for AlF. Despite HF showing the largest value of $B_0$, the interaction potential for HF-He is very isotropic, and hence $C_{6}^{1} $ is very small, yielding a low rotational quenching efficiency. Finally, NaH+He presents the same rotational quenching efficiency as AlF+He due to its smaller reduced mass and larger rotational constant, despite having a more isotropic PES than the AlF-He case.


\begin{table}
\caption{Parameters required for the evaluation of the scaling law for the rotational quenching} 
\begin{tabular}{l c c c} 
\hline\hline   

 Molecule & \hspace{2 pt} $B_{0}$(cm$^{-1}$) & \hspace{2
 pt} C$_{6}^{1}$ (cm$^{-1}$$\AA^{-6}$)  & \hspace{2
 pt} $\mu_{D}$ (a.m.u) 
\\ [0.5ex]  
\hline   
\\
AlF & \hspace{2pt} 0.54999 & \hspace{2pt} 92678.9 & \hspace{2 pt}  3.6820736 \\[1 ex]
CaF & \hspace{2pt} 0.33850 & \hspace{2pt} 71208.3 & \hspace{2 pt} 3.7486183 \\[1 ex]
SrF & \hspace{2pt} 0.25053 & \hspace{2pt} 31293.0 & \hspace{2 pt} 3.1862378  \\[1 ex]
HF & \hspace{2pt} 20.9557 & \hspace{2pt} 4537  & \hspace{2 pt} 3.36192483  \\[1 ex]
NaH & \hspace{2pt} 4.89 & \hspace{2pt} 39997.0 & \hspace{2 pt} 3.430440591 \\[1 ex]
\label{tab:table_BDWA}
\end{tabular}  
\end{table}

\subsection{Sensitivity to the accuracy of the PES}

In the cold regime, the accuracy of ab initio PESs may not be sufficient to guarantee the quality of the dynamics results. Therefore, we tested the role of the accuracy of the PES on the quantal dynamics, and the results are shown in Figure 7. This Figure shows the elastic cross section for CaF+He and SrF+He systems when the PES are shifted up and down by 2.5$\%$. As expected, in the ultracold regime ($E_k \lesssim1$mK), small changes in the interaction potential lead to drastic changes in the cross section. On the contrary, for $E_k \gtrsim$0.1~K, the elastic cross section shows the same trend as a function of the collision energy and the density of resonances is unaltered by the changes on the interaction potential, as previously shown in the AlF-He system~\cite{Sangami2024}.  


\begin{figure}[htbp]
\centering
\includegraphics[scale=0.32]{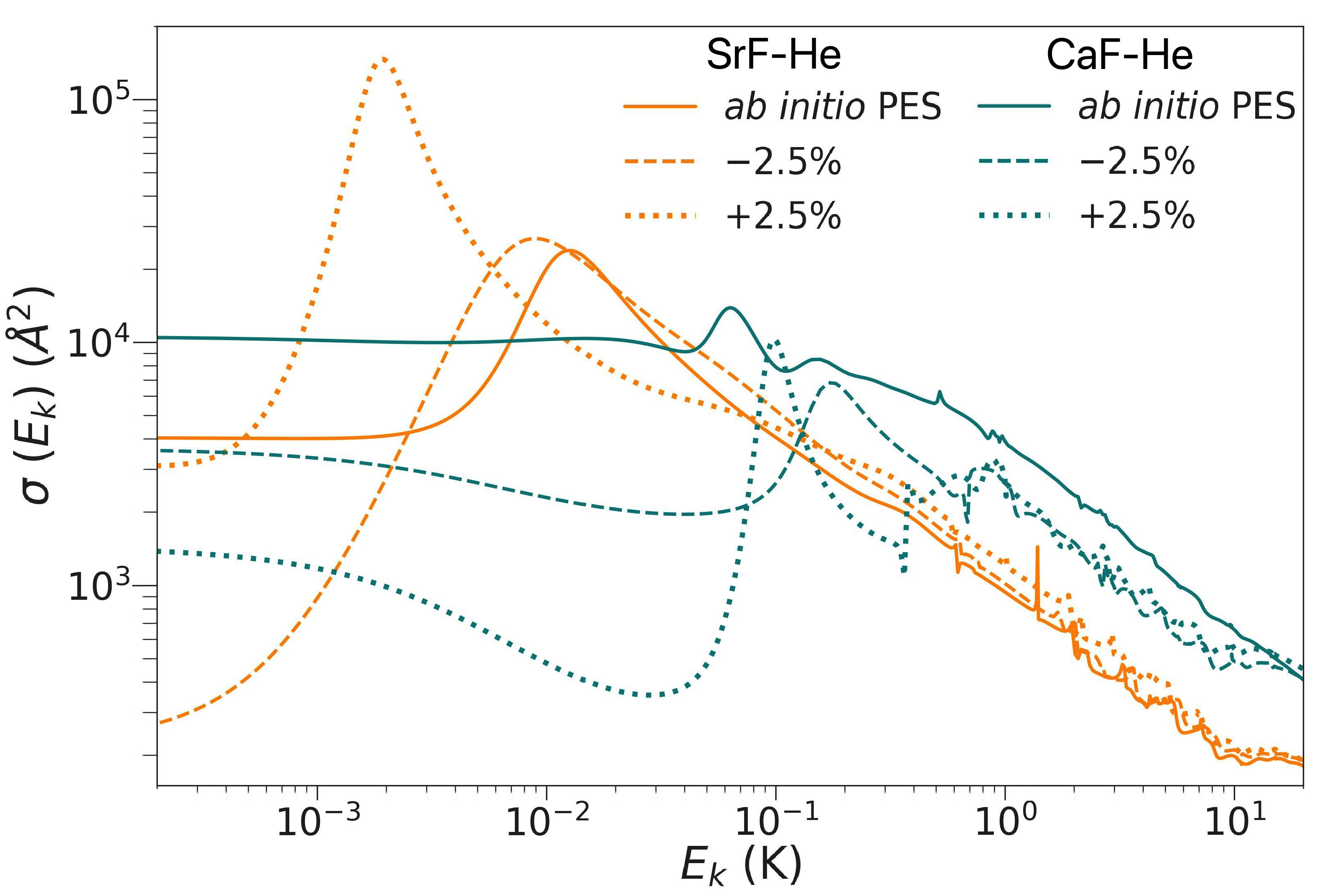}
\caption{Sensitivity of the elastic cross section to details of the PES. Solid lines represent the elastic cross section as a function fo the collision energy for the PES employed along this work. The dotted and dashed lines stand for the elastic cross section as a function of the collision energy when the PES is shifted by $\pm$2.5$\%$, respectively.  }
\label{fig:error}
\end{figure}

\subsection{Transport properties}
\label{sect:transport}

In this section, we focus on the elastic transport properties of XF-He, with X=H, Al, Ca, and Sr, relevant for buffer gas cooling experiments. First, we explore the thermalization efficiency of the translational degrees of freedom of molecules by the buffer gas given by the elastic-to-inelastic ratio of rate constants. The results for $k_{1}/k_{1\rightarrow0}$ and $k_{2}/(k_{2\rightarrow0} + k_{2\rightarrow0}) $ are displayed in Table \ref{tab:thermalization_rates}, showing that AlF does not thermalize efficiently due to a large rotational quenching efficiency as discussed above. On the contrary, CaF and Sr show a rather large value, meaning that many elastic collisions occur per inelastic event due to a less efficient rotational quenching, and the higher the temperature, the higher the thermalization rate. 

\begin{table}
\caption{Thermalization ratios of the $j = 1,2$ channels for the different molecular species. Three different temperatures are considered to cover a broad region of the buffer gas cooling experiment }
\begin{tabular}{@{\extracolsep{\fill}}lcc}

\textbf{Molecule} & \textbf{$k_{1}/k_{1\rightarrow0} $} & \textbf{$k_{2}/(k_{2\rightarrow1} +k_{2\rightarrow0} ) $ }  \\
\midrule
\hline
\vspace{2 px}
\\
&\hspace{25 px} \textbf{T = 0.5 K}& \hspace{15 px} \\
\midrule
\hline
\\
AlF + He & 2.07 & 2.58  \\
CaF + He & 53.1 & 13.90  \\
SrF + He & 109.2 & 7.28  \\
HF + He & 5400 & 58.4  \\
\\
\hline
\vspace{2 px}
\\
&\hspace{20 px} \textbf{T = 4 K}&\\
\midrule
\hline
\\
AlF + He & 7.68 &  5.01\\
CaF + He & 150.6 & 10.8  \\
SrF + He & 55.3 & 7.39  \\
HF + He & 1$\times$10$^{4}$ & 87.5  \\
\\
\hline
\vspace{2 px}
\\
&\hspace{20 px} \textbf{T = 10 K}&\\
\midrule
\hline
\\
AlF + He & 12.6 &  8.04 \\
CaF + He & 171 & 9.89  \\
SrF + He & 51.6 & 8.01   \\
HF + He & 1277 & 101 \\
\\
\bottomrule
\label{tab:thermalization_rates}
\end{tabular}
\end{table}

\begin{figure}[t] %
    \centering
    \includegraphics[width=\linewidth]{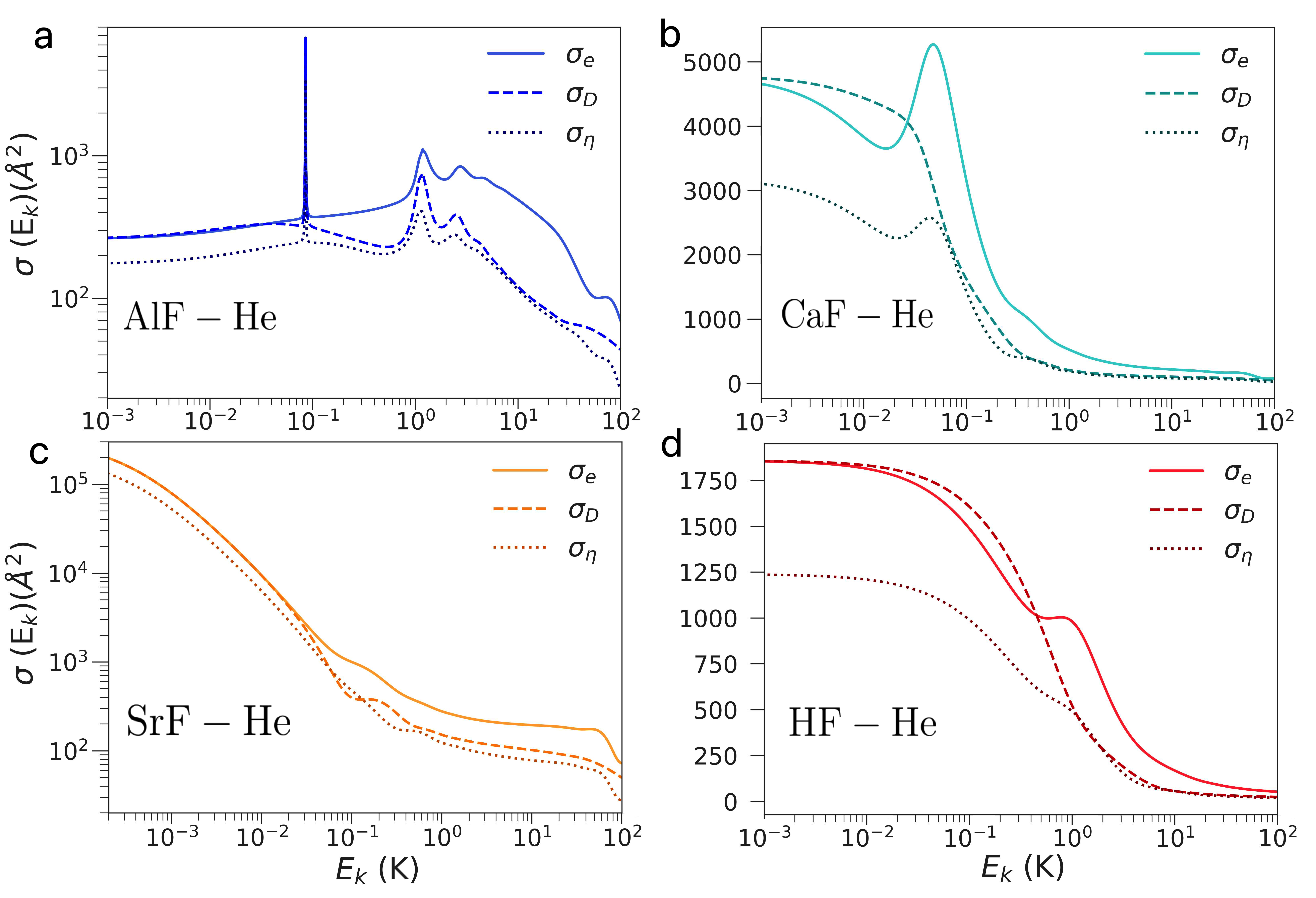} 
    \caption{Elastic ($\sigma_{e}$), Diffusive ($\sigma_{D}$) and viscosity ($\sigma_{\eta}$) cross sections as a function of the energy, for AlF-He (panel a), CaF-He (panel b), SrF-He (panel c) and HF-He (panel d). }
    \label{fig:sc_xs}
\end{figure}

Additionally, we have computed the transport cross section in the single channel formalism described by Eqs.~(\ref{eq:dxs_sc}) and (\ref{eq:vxs_sc}) and the elastic cross section considering $V_{0}(R)$ as the interaction potential. The results for all the monofluorides are shown in Figure \ref{fig:sc_xs}. In the cold regime, $E_k\lesssim 1$~K, SrF, and CaF-He show a large diffusive cross section, inducing more frequent collisions with the buffer gas atoms, which can hinder the transport of the cold molecules inside the buffer cell. All transport cross sections decay to lower values for larger collision energies, indicating a large diffusion coefficient for all monofluorides under consideration. In the case of CaF-He, we notice a broad shape resonance round 50~mK, showing in the elastic and viscosity cross sections. Based on the results from Fig.~\ref{fig:error}, the shape resonance will persist even if the present potential shows a 5$\%$ accuracy, although it may appear at a different collision energy.  

\label{fig:error}

\begin{figure}[htbp]
\centering
\includegraphics[scale=0.36]{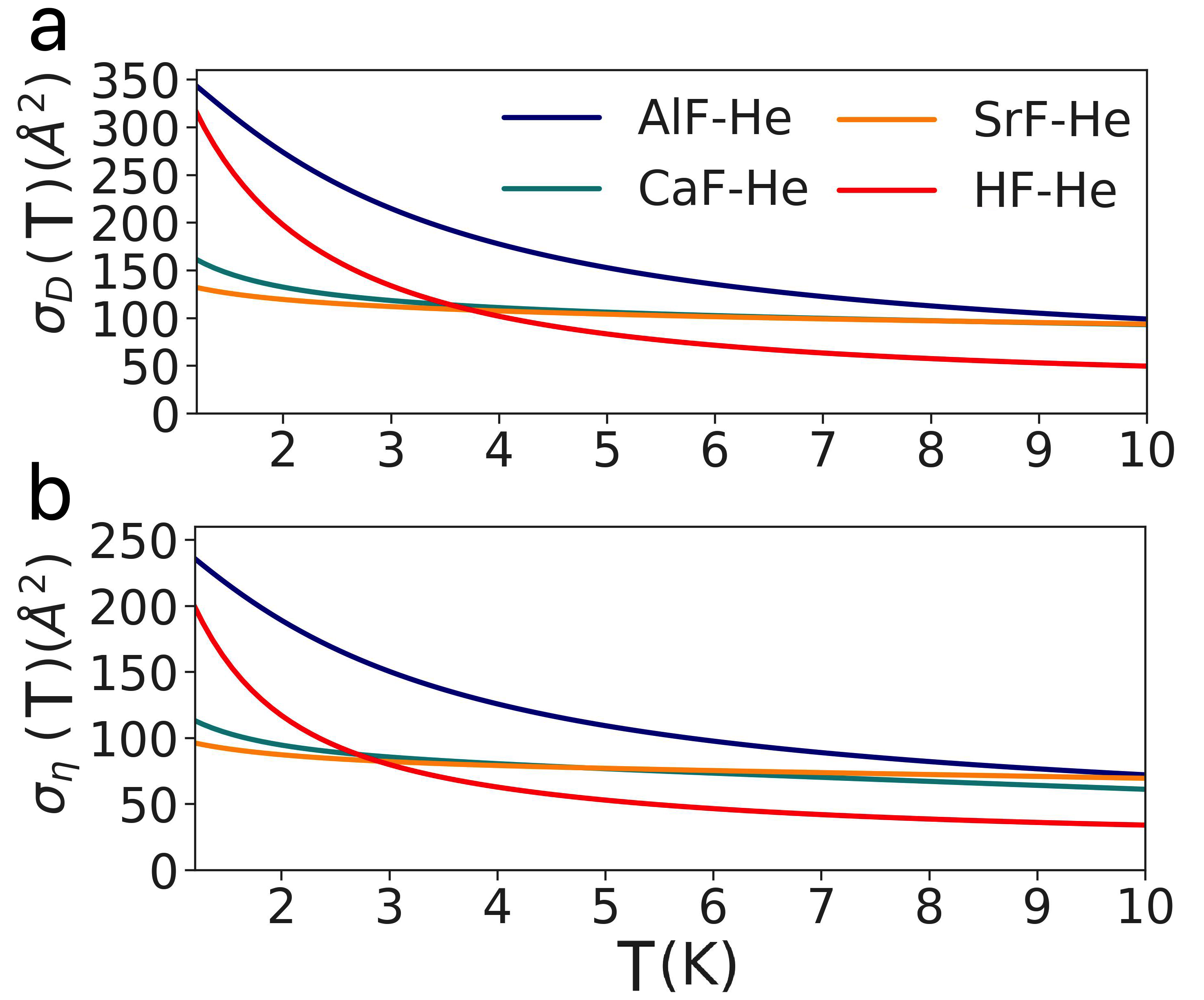}
\caption{Thermally averaged quantum transport cross sections as a function of temperature for all XF+He systems. Panel (a) presents the averaged diffusive cross section, while panel (b) displays the averaged viscosity cross section.}
\label{fig:avg_trans_xs}
\end{figure}

Next, we focus on the temperature range for buffer gas cooling experiments, and we calculate thermally averaged transport cross section, and the results are displayed in Figure~\ref{fig:avg_trans_xs}. We notice that AlF-He presents the largest thermally averaged diffusive and viscosity cross-section, which translates into small diffusion and viscosity coefficients. HF-He shows the opposite behavior: small thermal averaged cross sections and, hence, large diffusion and viscosity coefficients. At temperatures $T\gtrsim 8$~K, CaF, SrF and AlF-He show the same thermal averaged transport cross section, independent of the metal species and molecular symmetry, as suggested in \cite{Karra2022}. However, the exception to the rule is HF, which asymptotic value undergoes the previous one, suggesting that the universal asymptotic law for the diffusion, and even viscosity, cross section applies only to the relative strong ionic compounds Metal+F.



\section{conclusions}
\label{sect:conclusions}
We have studied the quantum dynamics of rotational and translational relaxation of monofluorides (XF), where X= Al, Ca, Sr, and H, in the presence of a cold helium gas, with a focus on buffer gas cooling. To achieve this, we computed the potential energy surfaces for the XF-He system within the rigid rotor approximation using ab initio quantum chemical calculations. These potentials were then employed to determine state-to-state scattering cross sections and rate constants through the numerical solution of the coupled-channel equations. The results were further analyzed using the Born Distorted Wave Approximation, providing insight into the state-to-state rate constants based on the molecular properties and interaction characteristics. The relationship derived explains why AlF shows the largest relaxation rate among all XF studied and, qualitatively, the hierarchy of the rotational quenching rate. Using this expression, we predict that MgF will show a rotational quenching rate similar to CaF. In contrast, BaF and RaF in a helium buffer gas will show a similar rotational quenching rate as SrF based on the similitude in the rotational constraints and anisotropy of the interaction. On the other hand, we have calculated the transport cross sections, which are relevant to understanding the buffer gas dynamics. AlF shows the largest diffusion and viscosity cross sections, whereas SrF and CaF show almost identical values. Therefore, AlF will show a smaller diffusion coefficient and viscosity than CaF and SrF in He.

Finally, from our global study on XF-He collisions, we have clarified the nature of the rotational quenching and reduced it to a straightforward relationship. However, it is still necessary to account for vibrational quenching and possible rotation-vibrational coupling since molecules after ablation could appear in highly vibrational states, which will be the focus of future work.

\section{Acknowledgments}
The authors thank Dr. Tim Langen for useful discussions and for bringing the topic of this work to our attention. Also, authors  thank Dr. Cheikh Bop for generously sharing the potential data for the NaH+He system. This work has been supported by the Simons Foundation.

\bibliography{main.bib}

\begin{thebibliography}{48}%
\makeatletter
\providecommand \@ifxundefined [1]{%
 \@ifx{#1\undefined}
}%
\providecommand \@ifnum [1]{%
 \ifnum #1\expandafter \@firstoftwo
 \else \expandafter \@secondoftwo
 \fi
}%
\providecommand \@ifx [1]{%
 \ifx #1\expandafter \@firstoftwo
 \else \expandafter \@secondoftwo
 \fi
}%
\providecommand \natexlab [1]{#1}%
\providecommand \enquote  [1]{``#1''}%
\providecommand \bibnamefont  [1]{#1}%
\providecommand \bibfnamefont [1]{#1}%
\providecommand \citenamefont [1]{#1}%
\providecommand \href@noop [0]{\@secondoftwo}%
\providecommand \href [0]{\begingroup \@sanitize@url \@href}%
\providecommand \@href[1]{\@@startlink{#1}\@@href}%
\providecommand \@@href[1]{\endgroup#1\@@endlink}%
\providecommand \@sanitize@url [0]{\catcode `\\12\catcode `\$12\catcode
  `\&12\catcode `\#12\catcode `\^12\catcode `\_12\catcode `\%12\relax}%
\providecommand \@@startlink[1]{}%
\providecommand \@@endlink[0]{}%
\providecommand \url  [0]{\begingroup\@sanitize@url \@url }%
\providecommand \@url [1]{\endgroup\@href {#1}{\urlprefix }}%
\providecommand \urlprefix  [0]{URL }%
\providecommand \Eprint [0]{\href }%
\providecommand \doibase [0]{https://doi.org/}%
\providecommand \selectlanguage [0]{\@gobble}%
\providecommand \bibinfo  [0]{\@secondoftwo}%
\providecommand \bibfield  [0]{\@secondoftwo}%
\providecommand \translation [1]{[#1]}%
\providecommand \BibitemOpen [0]{}%
\providecommand \bibitemStop [0]{}%
\providecommand \bibitemNoStop [0]{.\EOS\space}%
\providecommand \EOS [0]{\spacefactor3000\relax}%
\providecommand \BibitemShut  [1]{\csname bibitem#1\endcsname}%
\let\auto@bib@innerbib\@empty
\bibitem [{\citenamefont {DeMille}(2002)}]{DeMille2002}%
  \BibitemOpen
  \bibfield  {author} {\bibinfo {author} {\bibfnamefont {D.}~\bibnamefont
  {DeMille}},\ }\bibfield  {title} {\bibinfo {title} {Quantum computation with
  trapped polar molecules},\ }\href
  {https://doi.org/10.1103/PhysRevLett.88.067901} {\bibfield  {journal}
  {\bibinfo  {journal} {Phys. Rev. Lett.}\ }\textbf {\bibinfo {volume} {88}},\
  \bibinfo {pages} {067901} (\bibinfo {year} {2002})}\BibitemShut {NoStop}%
\bibitem [{\citenamefont {Bohn}\ \emph {et~al.}(2017)\citenamefont {Bohn},
  \citenamefont {Rey},\ and\ \citenamefont {Ye}}]{Bohn2017}%
  \BibitemOpen
  \bibfield  {author} {\bibinfo {author} {\bibfnamefont {J.~L.}\ \bibnamefont
  {Bohn}}, \bibinfo {author} {\bibfnamefont {A.~M.}\ \bibnamefont {Rey}},\ and\
  \bibinfo {author} {\bibfnamefont {J.}~\bibnamefont {Ye}},\ }\bibfield
  {title} {\bibinfo {title} {Cold molecules: Progress in quantum engineering of
  chemistry and quantum matter},\ }\href
  {https://api.semanticscholar.org/CorpusID:206656605} {\bibfield  {journal}
  {\bibinfo  {journal} {Science}\ }\textbf {\bibinfo {volume} {357}},\ \bibinfo
  {pages} {1002 } (\bibinfo {year} {2017})}\BibitemShut {NoStop}%
\bibitem [{\citenamefont {{Kaufman}}\ and\ \citenamefont
  {{Ni}}(2021)}]{Kaufman2021}%
  \BibitemOpen
  \bibfield  {author} {\bibinfo {author} {\bibfnamefont {A.~M.}\ \bibnamefont
  {{Kaufman}}}\ and\ \bibinfo {author} {\bibfnamefont {K.-K.}\ \bibnamefont
  {{Ni}}},\ }\bibfield  {title} {\bibinfo {title} {{Quantum science with
  optical tweezer arrays of ultracold atoms and molecules}},\ }\href
  {https://doi.org/10.1038/s41567-021-01357-2} {\bibfield  {journal} {\bibinfo
  {journal} {Nature Physics}\ }\textbf {\bibinfo {volume} {17}},\ \bibinfo
  {pages} {1324} (\bibinfo {year} {2021})}\BibitemShut {NoStop}%
\bibitem [{\citenamefont {{Cornish}}\ \emph {et~al.}(2024)\citenamefont
  {{Cornish}}, \citenamefont {{Tarbutt}},\ and\ \citenamefont
  {{Hazzard}}}]{Cornish2024}%
  \BibitemOpen
  \bibfield  {author} {\bibinfo {author} {\bibfnamefont {S.~L.}\ \bibnamefont
  {{Cornish}}}, \bibinfo {author} {\bibfnamefont {M.~R.}\ \bibnamefont
  {{Tarbutt}}},\ and\ \bibinfo {author} {\bibfnamefont {K.~R.~A.}\ \bibnamefont
  {{Hazzard}}},\ }\bibfield  {title} {\bibinfo {title} {{Quantum computation
  and quantum simulation with ultracold molecules}},\ }\href
  {https://doi.org/10.1038/s41567-024-02453-9} {\bibfield  {journal} {\bibinfo
  {journal} {Nature Physics}\ }\textbf {\bibinfo {volume} {20}},\ \bibinfo
  {pages} {730} (\bibinfo {year} {2024})},\ \Eprint
  {https://arxiv.org/abs/2401.05086} {arXiv:2401.05086 [cond-mat.quant-gas]}
  \BibitemShut {NoStop}%
\bibitem [{\citenamefont {Carroll}\ \emph {et~al.}(2024)\citenamefont
  {Carroll}, \citenamefont {Hirzler}, \citenamefont {Miller}, \citenamefont
  {Wellnitz}, \citenamefont {Muleady}, \citenamefont {Lin}, \citenamefont
  {Zamarski}, \citenamefont {Wang}, \citenamefont {Bohn}, \citenamefont {Rey},\
  and\ \citenamefont {Ye}}]{carroll2024}%
  \BibitemOpen
  \bibfield  {author} {\bibinfo {author} {\bibfnamefont {A.~N.}\ \bibnamefont
  {Carroll}}, \bibinfo {author} {\bibfnamefont {H.}~\bibnamefont {Hirzler}},
  \bibinfo {author} {\bibfnamefont {C.}~\bibnamefont {Miller}}, \bibinfo
  {author} {\bibfnamefont {D.}~\bibnamefont {Wellnitz}}, \bibinfo {author}
  {\bibfnamefont {S.~R.}\ \bibnamefont {Muleady}}, \bibinfo {author}
  {\bibfnamefont {J.}~\bibnamefont {Lin}}, \bibinfo {author} {\bibfnamefont
  {K.~P.}\ \bibnamefont {Zamarski}}, \bibinfo {author} {\bibfnamefont
  {R.~R.~W.}\ \bibnamefont {Wang}}, \bibinfo {author} {\bibfnamefont {J.~L.}\
  \bibnamefont {Bohn}}, \bibinfo {author} {\bibfnamefont {A.~M.}\ \bibnamefont
  {Rey}},\ and\ \bibinfo {author} {\bibfnamefont {J.}~\bibnamefont {Ye}},\
  }\href {https://arxiv.org/abs/2404.18916} {\bibinfo {title} {Observation of
  generalized t-j spin dynamics with tunable dipolar interactions}} (\bibinfo
  {year} {2024}),\ \Eprint {https://arxiv.org/abs/2404.18916} {arXiv:2404.18916
  [cond-mat.quant-gas]} \BibitemShut {NoStop}%
\bibitem [{\citenamefont {{Sch{\"a}fer}}\ \emph {et~al.}(2020)\citenamefont
  {{Sch{\"a}fer}}, \citenamefont {{Fukuhara}}, \citenamefont {{Sugawa}},
  \citenamefont {{Takasu}},\ and\ \citenamefont {{Takahashi}}}]{schafer2020}%
  \BibitemOpen
  \bibfield  {author} {\bibinfo {author} {\bibfnamefont {F.}~\bibnamefont
  {{Sch{\"a}fer}}}, \bibinfo {author} {\bibfnamefont {T.}~\bibnamefont
  {{Fukuhara}}}, \bibinfo {author} {\bibfnamefont {S.}~\bibnamefont
  {{Sugawa}}}, \bibinfo {author} {\bibfnamefont {Y.}~\bibnamefont {{Takasu}}},\
  and\ \bibinfo {author} {\bibfnamefont {Y.}~\bibnamefont {{Takahashi}}},\
  }\bibfield  {title} {\bibinfo {title} {{Tools for quantum simulation with
  ultracold atoms in optical lattices}},\ }\href
  {https://doi.org/10.1038/s42254-020-0195-3} {\bibfield  {journal} {\bibinfo
  {journal} {Nature Reviews Physics}\ }\textbf {\bibinfo {volume} {2}},\
  \bibinfo {pages} {411} (\bibinfo {year} {2020})},\ \Eprint
  {https://arxiv.org/abs/2006.06120} {arXiv:2006.06120 [cond-mat.quant-gas]}
  \BibitemShut {NoStop}%
\bibitem [{\citenamefont {Langen}(2022)}]{Langen2022}%
  \BibitemOpen
  \bibfield  {author} {\bibinfo {author} {\bibfnamefont {T.}~\bibnamefont
  {Langen}},\ }\bibfield  {title} {\bibinfo {title} {Dipolar supersolids: Solid
  and superfluid at the same time},\ }\href {https://doi.org/10.1063/PT.3.4961}
  {\bibfield  {journal} {\bibinfo  {journal} {Physics Today}\ }\textbf
  {\bibinfo {volume} {75}},\ \bibinfo {pages} {36} (\bibinfo {year} {2022})},\
  \Eprint
  {https://arxiv.org/abs/https://pubs.aip.org/physicstoday/article-pdf/75/3/36/16334017/36\_1\_online.pdf}
  {https://pubs.aip.org/physicstoday/article-pdf/75/3/36/16334017/36\_1\_online.pdf}
  \BibitemShut {NoStop}%
\bibitem [{\citenamefont {Schmidt}\ \emph {et~al.}(2022)\citenamefont
  {Schmidt}, \citenamefont {Lassabli\`ere}, \citenamefont {Qu\'em\'ener},\ and\
  \citenamefont {Langen}}]{Langen2022b}%
  \BibitemOpen
  \bibfield  {author} {\bibinfo {author} {\bibfnamefont {M.}~\bibnamefont
  {Schmidt}}, \bibinfo {author} {\bibfnamefont {L.}~\bibnamefont
  {Lassabli\`ere}}, \bibinfo {author} {\bibfnamefont {G.}~\bibnamefont
  {Qu\'em\'ener}},\ and\ \bibinfo {author} {\bibfnamefont {T.}~\bibnamefont
  {Langen}},\ }\bibfield  {title} {\bibinfo {title} {Self-bound dipolar
  droplets and supersolids in molecular bose-einstein condensates},\ }\href
  {https://doi.org/10.1103/PhysRevResearch.4.013235} {\bibfield  {journal}
  {\bibinfo  {journal} {Phys. Rev. Res.}\ }\textbf {\bibinfo {volume} {4}},\
  \bibinfo {pages} {013235} (\bibinfo {year} {2022})}\BibitemShut {NoStop}%
\bibitem [{\citenamefont {Karman}\ \emph {et~al.}(2024)\citenamefont {Karman},
  \citenamefont {Tomza},\ and\ \citenamefont {P\'erez-R\'\i{}os}}]{Karman2024}%
  \BibitemOpen
  \bibfield  {author} {\bibinfo {author} {\bibfnamefont {T.}~\bibnamefont
  {Karman}}, \bibinfo {author} {\bibfnamefont {M.}~\bibnamefont {Tomza}},\ and\
  \bibinfo {author} {\bibfnamefont {J.}~\bibnamefont {P\'erez-R\'\i{}os}},\
  }\bibfield  {title} {\bibinfo {title} {{Ultracold chemistry as a testbed for
  few-body physics}},\ }\href {https://doi.org/10.1038/s41567-024-02467-3}
  {\bibfield  {journal} {\bibinfo  {journal} {Nature Phys.}\ }\textbf {\bibinfo
  {volume} {20}},\ \bibinfo {pages} {722} (\bibinfo {year} {2024})}\BibitemShut
  {NoStop}%
\bibitem [{\citenamefont {Liu}\ \emph {et~al.}(2023)\citenamefont {Liu},
  \citenamefont {Zhu}, \citenamefont {Luke}, \citenamefont {Houwman},
  \citenamefont {Babin}, \citenamefont {Hu},\ and\ \citenamefont
  {Ni}}]{Liu2023}%
  \BibitemOpen
  \bibfield  {author} {\bibinfo {author} {\bibfnamefont {Y.-X.}\ \bibnamefont
  {Liu}}, \bibinfo {author} {\bibfnamefont {L.}~\bibnamefont {Zhu}}, \bibinfo
  {author} {\bibfnamefont {J.}~\bibnamefont {Luke}}, \bibinfo {author}
  {\bibfnamefont {J.~J.~A.}\ \bibnamefont {Houwman}}, \bibinfo {author}
  {\bibfnamefont {M.~C.}\ \bibnamefont {Babin}}, \bibinfo {author}
  {\bibfnamefont {M.-G.}\ \bibnamefont {Hu}},\ and\ \bibinfo {author}
  {\bibfnamefont {K.-K.}\ \bibnamefont {Ni}},\ }\href
  {https://arxiv.org/abs/2310.07620} {\bibinfo {title} {Quantum interference
  and entanglement in ultracold atom-exchange reactions}} (\bibinfo {year}
  {2023}),\ \Eprint {https://arxiv.org/abs/2310.07620} {arXiv:2310.07620
  [physics.atom-ph]} \BibitemShut {NoStop}%
\bibitem [{\citenamefont {Morita}\ \emph {et~al.}(2024)\citenamefont {Morita},
  \citenamefont {Kosicki}, \citenamefont {\ifmmode~\dot{Z}\else
  \.{Z}\fi{}uchowski}, \citenamefont {Brumer},\ and\ \citenamefont
  {Tscherbul}}]{Masato2024}%
  \BibitemOpen
  \bibfield  {author} {\bibinfo {author} {\bibfnamefont {M.}~\bibnamefont
  {Morita}}, \bibinfo {author} {\bibfnamefont {M.~B.}\ \bibnamefont {Kosicki}},
  \bibinfo {author} {\bibfnamefont {P.~S.}\ \bibnamefont {\ifmmode~\dot{Z}\else
  \.{Z}\fi{}uchowski}}, \bibinfo {author} {\bibfnamefont {P.}~\bibnamefont
  {Brumer}},\ and\ \bibinfo {author} {\bibfnamefont {T.~V.}\ \bibnamefont
  {Tscherbul}},\ }\bibfield  {title} {\bibinfo {title} {Magnetic feshbach
  resonances in ultracold atom-molecule collisions},\ }\href
  {https://doi.org/10.1103/PhysRevA.110.L021301} {\bibfield  {journal}
  {\bibinfo  {journal} {Phys. Rev. A}\ }\textbf {\bibinfo {volume} {110}},\
  \bibinfo {pages} {L021301} (\bibinfo {year} {2024})}\BibitemShut {NoStop}%
\bibitem [{\citenamefont {Lett}\ \emph {et~al.}(1995)\citenamefont {Lett},
  \citenamefont {Julienne},\ and\ \citenamefont {Phillips}}]{Lett1995}%
  \BibitemOpen
  \bibfield  {author} {\bibinfo {author} {\bibfnamefont {P.~D.}\ \bibnamefont
  {Lett}}, \bibinfo {author} {\bibfnamefont {P.~S.}\ \bibnamefont {Julienne}},\
  and\ \bibinfo {author} {\bibfnamefont {W.~D.}\ \bibnamefont {Phillips}},\
  }\bibfield  {title} {\bibinfo {title} {Photoassociative spectroscopy of
  laser-cooled atoms},\ }\href
  {https://doi.org/10.1146/annurev.pc.46.100195.002231} {\bibfield  {journal}
  {\bibinfo  {journal} {Annual Review of Physical Chemistry}\ }\textbf
  {\bibinfo {volume} {46}},\ \bibinfo {pages} {423} (\bibinfo {year} {1995})},\
  \bibinfo {note} {pMID: 24329894},\ \Eprint
  {https://arxiv.org/abs/https://doi.org/10.1146/annurev.pc.46.100195.002231}
  {https://doi.org/10.1146/annurev.pc.46.100195.002231} \BibitemShut {NoStop}%
\bibitem [{\citenamefont {Jones}\ \emph {et~al.}(2006)\citenamefont {Jones},
  \citenamefont {Tiesinga}, \citenamefont {Lett},\ and\ \citenamefont
  {Julienne}}]{PA}%
  \BibitemOpen
  \bibfield  {author} {\bibinfo {author} {\bibfnamefont {K.~M.}\ \bibnamefont
  {Jones}}, \bibinfo {author} {\bibfnamefont {E.}~\bibnamefont {Tiesinga}},
  \bibinfo {author} {\bibfnamefont {P.~D.}\ \bibnamefont {Lett}},\ and\
  \bibinfo {author} {\bibfnamefont {P.~S.}\ \bibnamefont {Julienne}},\
  }\bibfield  {title} {\bibinfo {title} {Ultracold photoassociation
  spectroscopy: Long-range molecules and atomic scattering},\ }\href
  {https://doi.org/10.1103/RevModPhys.78.483} {\bibfield  {journal} {\bibinfo
  {journal} {Rev. Mod. Phys.}\ }\textbf {\bibinfo {volume} {78}},\ \bibinfo
  {pages} {483} (\bibinfo {year} {2006})}\BibitemShut {NoStop}%
\bibitem [{\citenamefont {Fioretti}\ \emph {et~al.}(1998)\citenamefont
  {Fioretti}, \citenamefont {Comparat}, \citenamefont {Crubellier},
  \citenamefont {Dulieu}, \citenamefont {Masnou-Seeuws},\ and\ \citenamefont
  {Pillet}}]{PA2}%
  \BibitemOpen
  \bibfield  {author} {\bibinfo {author} {\bibfnamefont {A.}~\bibnamefont
  {Fioretti}}, \bibinfo {author} {\bibfnamefont {D.}~\bibnamefont {Comparat}},
  \bibinfo {author} {\bibfnamefont {A.}~\bibnamefont {Crubellier}}, \bibinfo
  {author} {\bibfnamefont {O.}~\bibnamefont {Dulieu}}, \bibinfo {author}
  {\bibfnamefont {F.}~\bibnamefont {Masnou-Seeuws}},\ and\ \bibinfo {author}
  {\bibfnamefont {P.}~\bibnamefont {Pillet}},\ }\bibfield  {title} {\bibinfo
  {title} {Formation of cold ${\mathrm{cs}}_{2}$ molecules through
  photoassociation},\ }\href {https://doi.org/10.1103/PhysRevLett.80.4402}
  {\bibfield  {journal} {\bibinfo  {journal} {Phys. Rev. Lett.}\ }\textbf
  {\bibinfo {volume} {80}},\ \bibinfo {pages} {4402} (\bibinfo {year}
  {1998})}\BibitemShut {NoStop}%
\bibitem [{\citenamefont {P\'erez-R\'{\i}os}\ \emph {et~al.}(2015)\citenamefont
  {P\'erez-R\'{\i}os}, \citenamefont {Lepers},\ and\ \citenamefont
  {Dulieu}}]{PAJPR}%
  \BibitemOpen
  \bibfield  {author} {\bibinfo {author} {\bibfnamefont {J.}~\bibnamefont
  {P\'erez-R\'{\i}os}}, \bibinfo {author} {\bibfnamefont {M.}~\bibnamefont
  {Lepers}},\ and\ \bibinfo {author} {\bibfnamefont {O.}~\bibnamefont
  {Dulieu}},\ }\bibfield  {title} {\bibinfo {title} {Theory of long-range
  ultracold atom-molecule photoassociation},\ }\href
  {https://doi.org/10.1103/PhysRevLett.115.073201} {\bibfield  {journal}
  {\bibinfo  {journal} {Phys. Rev. Lett.}\ }\textbf {\bibinfo {volume} {115}},\
  \bibinfo {pages} {073201} (\bibinfo {year} {2015})}\BibitemShut {NoStop}%
\bibitem [{\citenamefont {Moerdijk}\ \emph {et~al.}(1995)\citenamefont
  {Moerdijk}, \citenamefont {Verhaar},\ and\ \citenamefont
  {Axelsson}}]{Moerdijk}%
  \BibitemOpen
  \bibfield  {author} {\bibinfo {author} {\bibfnamefont {A.~J.}\ \bibnamefont
  {Moerdijk}}, \bibinfo {author} {\bibfnamefont {B.~J.}\ \bibnamefont
  {Verhaar}},\ and\ \bibinfo {author} {\bibfnamefont {A.}~\bibnamefont
  {Axelsson}},\ }\bibfield  {title} {\bibinfo {title} {Resonances in ultracold
  collisions of $^{6}\mathrm{Li}$, $^{7}\mathrm{Li}$, and $^{23}\mathrm{Na}$},\
  }\href {https://doi.org/10.1103/PhysRevA.51.4852} {\bibfield  {journal}
  {\bibinfo  {journal} {Phys. Rev. A}\ }\textbf {\bibinfo {volume} {51}},\
  \bibinfo {pages} {4852} (\bibinfo {year} {1995})}\BibitemShut {NoStop}%
\bibitem [{\citenamefont {Chin}\ \emph {et~al.}(2010)\citenamefont {Chin},
  \citenamefont {Grimm}, \citenamefont {Julienne},\ and\ \citenamefont
  {Tiesinga}}]{Magnetoassociation}%
  \BibitemOpen
  \bibfield  {author} {\bibinfo {author} {\bibfnamefont {C.}~\bibnamefont
  {Chin}}, \bibinfo {author} {\bibfnamefont {R.}~\bibnamefont {Grimm}},
  \bibinfo {author} {\bibfnamefont {P.}~\bibnamefont {Julienne}},\ and\
  \bibinfo {author} {\bibfnamefont {E.}~\bibnamefont {Tiesinga}},\ }\bibfield
  {title} {\bibinfo {title} {Feshbach resonances in ultracold gases},\ }\href
  {https://doi.org/10.1103/RevModPhys.82.1225} {\bibfield  {journal} {\bibinfo
  {journal} {Rev. Mod. Phys.}\ }\textbf {\bibinfo {volume} {82}},\ \bibinfo
  {pages} {1225} (\bibinfo {year} {2010})}\BibitemShut {NoStop}%
\bibitem [{\citenamefont {Ruttley}\ \emph {et~al.}(2023)\citenamefont
  {Ruttley}, \citenamefont {Guttridge}, \citenamefont {Spence}, \citenamefont
  {Bird}, \citenamefont {Le~Sueur}, \citenamefont {Hutson},\ and\ \citenamefont
  {Cornish}}]{MergeAssociation}%
  \BibitemOpen
  \bibfield  {author} {\bibinfo {author} {\bibfnamefont {D.~K.}\ \bibnamefont
  {Ruttley}}, \bibinfo {author} {\bibfnamefont {A.}~\bibnamefont {Guttridge}},
  \bibinfo {author} {\bibfnamefont {S.}~\bibnamefont {Spence}}, \bibinfo
  {author} {\bibfnamefont {R.~C.}\ \bibnamefont {Bird}}, \bibinfo {author}
  {\bibfnamefont {C.~R.}\ \bibnamefont {Le~Sueur}}, \bibinfo {author}
  {\bibfnamefont {J.~M.}\ \bibnamefont {Hutson}},\ and\ \bibinfo {author}
  {\bibfnamefont {S.~L.}\ \bibnamefont {Cornish}},\ }\bibfield  {title}
  {\bibinfo {title} {Formation of ultracold molecules by merging optical
  tweezers},\ }\href {https://doi.org/10.1103/PhysRevLett.130.223401}
  {\bibfield  {journal} {\bibinfo  {journal} {Phys. Rev. Lett.}\ }\textbf
  {\bibinfo {volume} {130}},\ \bibinfo {pages} {223401} (\bibinfo {year}
  {2023})}\BibitemShut {NoStop}%
\bibitem [{\citenamefont {Bethlem}\ \emph {et~al.}(1999)\citenamefont
  {Bethlem}, \citenamefont {Berden},\ and\ \citenamefont {Meijer}}]{Stark}%
  \BibitemOpen
  \bibfield  {author} {\bibinfo {author} {\bibfnamefont {H.~L.}\ \bibnamefont
  {Bethlem}}, \bibinfo {author} {\bibfnamefont {G.}~\bibnamefont {Berden}},\
  and\ \bibinfo {author} {\bibfnamefont {G.}~\bibnamefont {Meijer}},\
  }\bibfield  {title} {\bibinfo {title} {Decelerating neutral dipolar
  molecules},\ }\href {https://doi.org/10.1103/PhysRevLett.83.1558} {\bibfield
  {journal} {\bibinfo  {journal} {Phys. Rev. Lett.}\ }\textbf {\bibinfo
  {volume} {83}},\ \bibinfo {pages} {1558} (\bibinfo {year}
  {1999})}\BibitemShut {NoStop}%
\bibitem [{\citenamefont {Vanhaecke}\ \emph {et~al.}(2007)\citenamefont
  {Vanhaecke}, \citenamefont {Meier}, \citenamefont {Andrist}, \citenamefont
  {Meier},\ and\ \citenamefont {Merkt}}]{Zeeman}%
  \BibitemOpen
  \bibfield  {author} {\bibinfo {author} {\bibfnamefont {N.}~\bibnamefont
  {Vanhaecke}}, \bibinfo {author} {\bibfnamefont {U.}~\bibnamefont {Meier}},
  \bibinfo {author} {\bibfnamefont {M.}~\bibnamefont {Andrist}}, \bibinfo
  {author} {\bibfnamefont {B.~H.}\ \bibnamefont {Meier}},\ and\ \bibinfo
  {author} {\bibfnamefont {F.}~\bibnamefont {Merkt}},\ }\bibfield  {title}
  {\bibinfo {title} {Multistage zeeman deceleration of hydrogen atoms},\ }\href
  {https://doi.org/10.1103/PhysRevA.75.031402} {\bibfield  {journal} {\bibinfo
  {journal} {Phys. Rev. A}\ }\textbf {\bibinfo {volume} {75}},\ \bibinfo
  {pages} {031402} (\bibinfo {year} {2007})}\BibitemShut {NoStop}%
\bibitem [{\citenamefont {Chervenkov}\ \emph {et~al.}(2014)\citenamefont
  {Chervenkov}, \citenamefont {Wu}, \citenamefont {Bayerl}, \citenamefont
  {Rohlfes}, \citenamefont {Gantner}, \citenamefont {Zeppenfeld},\ and\
  \citenamefont {Rempe}}]{Cherenkov}%
  \BibitemOpen
  \bibfield  {author} {\bibinfo {author} {\bibfnamefont {S.}~\bibnamefont
  {Chervenkov}}, \bibinfo {author} {\bibfnamefont {X.}~\bibnamefont {Wu}},
  \bibinfo {author} {\bibfnamefont {J.}~\bibnamefont {Bayerl}}, \bibinfo
  {author} {\bibfnamefont {A.}~\bibnamefont {Rohlfes}}, \bibinfo {author}
  {\bibfnamefont {T.}~\bibnamefont {Gantner}}, \bibinfo {author} {\bibfnamefont
  {M.}~\bibnamefont {Zeppenfeld}},\ and\ \bibinfo {author} {\bibfnamefont
  {G.}~\bibnamefont {Rempe}},\ }\bibfield  {title} {\bibinfo {title}
  {Continuous centrifuge decelerator for polar molecules},\ }\href
  {https://doi.org/10.1103/PhysRevLett.112.013001} {\bibfield  {journal}
  {\bibinfo  {journal} {Phys. Rev. Lett.}\ }\textbf {\bibinfo {volume} {112}},\
  \bibinfo {pages} {013001} (\bibinfo {year} {2014})}\BibitemShut {NoStop}%
\bibitem [{\citenamefont {Weinstein}\ \emph {et~al.}(1998)\citenamefont
  {Weinstein}, \citenamefont {DeCarvalho}, \citenamefont {Guillet},
  \citenamefont {Friedrich},\ and\ \citenamefont {Doyle}}]{Weinstein}%
  \BibitemOpen
  \bibfield  {author} {\bibinfo {author} {\bibfnamefont {J.~D.}\ \bibnamefont
  {Weinstein}}, \bibinfo {author} {\bibfnamefont {R.}~\bibnamefont
  {DeCarvalho}}, \bibinfo {author} {\bibfnamefont {T.}~\bibnamefont {Guillet}},
  \bibinfo {author} {\bibfnamefont {B.}~\bibnamefont {Friedrich}},\ and\
  \bibinfo {author} {\bibfnamefont {J.~M.}\ \bibnamefont {Doyle}},\ }\bibfield
  {title} {\bibinfo {title} {Magnetic trapping of calcium monohydride molecules
  at millikelvin temperatures},\ }\href@noop {} {\bibfield  {journal} {\bibinfo
   {journal} {Nature}\ }\textbf {\bibinfo {volume} {395}},\ \bibinfo {pages}
  {148} (\bibinfo {year} {1998})}\BibitemShut {NoStop}%
\bibitem [{\citenamefont {Hutzler}\ \emph {et~al.}(2012)\citenamefont
  {Hutzler}, \citenamefont {Lu},\ and\ \citenamefont {Doyle}}]{Buffergas1}%
  \BibitemOpen
  \bibfield  {author} {\bibinfo {author} {\bibfnamefont {N.~R.}\ \bibnamefont
  {Hutzler}}, \bibinfo {author} {\bibfnamefont {H.-I.}\ \bibnamefont {Lu}},\
  and\ \bibinfo {author} {\bibfnamefont {J.~M.}\ \bibnamefont {Doyle}},\
  }\bibfield  {title} {\bibinfo {title} {The buffer gas beam: An intense, cold,
  and slow source for atoms and molecules},\ }\bibfield  {booktitle} {\emph
  {\bibinfo {booktitle} {Chemical Reviews}},\ }\href
  {https://doi.org/10.1021/cr200362u} {\bibfield  {journal} {\bibinfo
  {journal} {Chemical Reviews}\ }\textbf {\bibinfo {volume} {112}},\ \bibinfo
  {pages} {4803} (\bibinfo {year} {2012})}\BibitemShut {NoStop}%
\bibitem [{\citenamefont {Truppe}\ \emph
  {et~al.}(2018{\natexlab{a}})\citenamefont {Truppe}, \citenamefont {Hambach},
  \citenamefont {Skoff}, \citenamefont {Bulleid}, \citenamefont {Bumby},
  \citenamefont {Hendricks}, \citenamefont {Hinds}, \citenamefont {Sauer},\
  and\ \citenamefont {Tarbutt}}]{Buffergas2}%
  \BibitemOpen
  \bibfield  {author} {\bibinfo {author} {\bibfnamefont {S.}~\bibnamefont
  {Truppe}}, \bibinfo {author} {\bibfnamefont {M.}~\bibnamefont {Hambach}},
  \bibinfo {author} {\bibfnamefont {S.~M.}\ \bibnamefont {Skoff}}, \bibinfo
  {author} {\bibfnamefont {N.~E.}\ \bibnamefont {Bulleid}}, \bibinfo {author}
  {\bibfnamefont {J.~S.}\ \bibnamefont {Bumby}}, \bibinfo {author}
  {\bibfnamefont {R.~J.}\ \bibnamefont {Hendricks}}, \bibinfo {author}
  {\bibfnamefont {E.~A.}\ \bibnamefont {Hinds}}, \bibinfo {author}
  {\bibfnamefont {B.~E.}\ \bibnamefont {Sauer}},\ and\ \bibinfo {author}
  {\bibfnamefont {M.~R.}\ \bibnamefont {Tarbutt}},\ }\bibfield  {title}
  {\bibinfo {title} {A buffer gas beam source for short, intense and slow
  molecular pulses},\ }\href {https://doi.org/10.1080/09500340.2017.1384516}
  {\bibfield  {journal} {\bibinfo  {journal} {Journal of Modern Optics}\
  }\textbf {\bibinfo {volume} {65}},\ \bibinfo {pages} {648} (\bibinfo {year}
  {2018}{\natexlab{a}})},\ \Eprint
  {https://arxiv.org/abs/https://doi.org/10.1080/09500340.2017.1384516}
  {https://doi.org/10.1080/09500340.2017.1384516} \BibitemShut {NoStop}%
\bibitem [{\citenamefont {Lim}\ \emph {et~al.}(2015)\citenamefont {Lim},
  \citenamefont {Frye}, \citenamefont {Hutson},\ and\ \citenamefont
  {Tarbutt}}]{sympa1}%
  \BibitemOpen
  \bibfield  {author} {\bibinfo {author} {\bibfnamefont {J.}~\bibnamefont
  {Lim}}, \bibinfo {author} {\bibfnamefont {M.~D.}\ \bibnamefont {Frye}},
  \bibinfo {author} {\bibfnamefont {J.~M.}\ \bibnamefont {Hutson}},\ and\
  \bibinfo {author} {\bibfnamefont {M.~R.}\ \bibnamefont {Tarbutt}},\
  }\bibfield  {title} {\bibinfo {title} {Modeling sympathetic cooling of
  molecules by ultracold atoms},\ }\href
  {https://doi.org/10.1103/PhysRevA.92.053419} {\bibfield  {journal} {\bibinfo
  {journal} {Phys. Rev. A}\ }\textbf {\bibinfo {volume} {92}},\ \bibinfo
  {pages} {053419} (\bibinfo {year} {2015})}\BibitemShut {NoStop}%
\bibitem [{\citenamefont {Son}\ \emph {et~al.}(2020)\citenamefont {Son},
  \citenamefont {Park}, \citenamefont {Ketterle},\ and\ \citenamefont
  {Jamison}}]{sympa2}%
  \BibitemOpen
  \bibfield  {author} {\bibinfo {author} {\bibfnamefont {H.}~\bibnamefont
  {Son}}, \bibinfo {author} {\bibfnamefont {J.~J.}\ \bibnamefont {Park}},
  \bibinfo {author} {\bibfnamefont {W.}~\bibnamefont {Ketterle}},\ and\
  \bibinfo {author} {\bibfnamefont {A.~O.}\ \bibnamefont {Jamison}},\
  }\bibfield  {title} {\bibinfo {title} {Collisional cooling of ultracold
  molecules},\ }\href {https://doi.org/10.1038/s41586-020-2141-z} {\bibfield
  {journal} {\bibinfo  {journal} {Nature}\ }\textbf {\bibinfo {volume} {580}},\
  \bibinfo {pages} {197} (\bibinfo {year} {2020})}\BibitemShut {NoStop}%
\bibitem [{\citenamefont {Tobias}\ \emph {et~al.}(2020)\citenamefont {Tobias},
  \citenamefont {Matsuda}, \citenamefont {Valtolina}, \citenamefont {De~Marco},
  \citenamefont {Li},\ and\ \citenamefont {Ye}}]{Tobias}%
  \BibitemOpen
  \bibfield  {author} {\bibinfo {author} {\bibfnamefont {W.~G.}\ \bibnamefont
  {Tobias}}, \bibinfo {author} {\bibfnamefont {K.}~\bibnamefont {Matsuda}},
  \bibinfo {author} {\bibfnamefont {G.}~\bibnamefont {Valtolina}}, \bibinfo
  {author} {\bibfnamefont {L.}~\bibnamefont {De~Marco}}, \bibinfo {author}
  {\bibfnamefont {J.-R.}\ \bibnamefont {Li}},\ and\ \bibinfo {author}
  {\bibfnamefont {J.}~\bibnamefont {Ye}},\ }\bibfield  {title} {\bibinfo
  {title} {Thermalization and sub-poissonian density fluctuations in a
  degenerate molecular fermi gas},\ }\href
  {https://doi.org/10.1103/PhysRevLett.124.033401} {\bibfield  {journal}
  {\bibinfo  {journal} {Phys. Rev. Lett.}\ }\textbf {\bibinfo {volume} {124}},\
  \bibinfo {pages} {033401} (\bibinfo {year} {2020})}\BibitemShut {NoStop}%
\bibitem [{\citenamefont {Shuman}\ \emph {et~al.}(2010)\citenamefont {Shuman},
  \citenamefont {Barry},\ and\ \citenamefont {DeMille}}]{lasercooling}%
  \BibitemOpen
  \bibfield  {author} {\bibinfo {author} {\bibfnamefont {E.~S.}\ \bibnamefont
  {Shuman}}, \bibinfo {author} {\bibfnamefont {J.~F.}\ \bibnamefont {Barry}},\
  and\ \bibinfo {author} {\bibfnamefont {D.}~\bibnamefont {DeMille}},\
  }\bibfield  {title} {\bibinfo {title} {Laser cooling of a diatomic
  molecule},\ }\href {https://doi.org/10.1038/nature09443} {\bibfield
  {journal} {\bibinfo  {journal} {Nature}\ }\textbf {\bibinfo {volume} {467}},\
  \bibinfo {pages} {820} (\bibinfo {year} {2010})}\BibitemShut {NoStop}%
\bibitem [{\citenamefont {Prehn}\ \emph {et~al.}(2016)\citenamefont {Prehn},
  \citenamefont {Ibr\"ugger}, \citenamefont {Gl\"ockner}, \citenamefont
  {Rempe},\ and\ \citenamefont {Zeppenfeld}}]{Prehn2016}%
  \BibitemOpen
  \bibfield  {author} {\bibinfo {author} {\bibfnamefont {A.}~\bibnamefont
  {Prehn}}, \bibinfo {author} {\bibfnamefont {M.}~\bibnamefont {Ibr\"ugger}},
  \bibinfo {author} {\bibfnamefont {R.}~\bibnamefont {Gl\"ockner}}, \bibinfo
  {author} {\bibfnamefont {G.}~\bibnamefont {Rempe}},\ and\ \bibinfo {author}
  {\bibfnamefont {M.}~\bibnamefont {Zeppenfeld}},\ }\bibfield  {title}
  {\bibinfo {title} {Optoelectrical cooling of polar molecules to
  submillikelvin temperatures},\ }\href
  {https://doi.org/10.1103/PhysRevLett.116.063005} {\bibfield  {journal}
  {\bibinfo  {journal} {Phys. Rev. Lett.}\ }\textbf {\bibinfo {volume} {116}},\
  \bibinfo {pages} {063005} (\bibinfo {year} {2016})}\BibitemShut {NoStop}%
\bibitem [{\citenamefont {Hemmerling}\ \emph {et~al.}(2016)\citenamefont
  {Hemmerling}, \citenamefont {Chae}, \citenamefont {Anderegg}, \citenamefont
  {Drayna}, \citenamefont {Hutzler}, \citenamefont {Collopy}, \citenamefont
  {Ye}, \citenamefont {Ketterle},\ and\ \citenamefont {Doyle}}]{Doyle2016}%
  \BibitemOpen
  \bibfield  {author} {\bibinfo {author} {\bibfnamefont {B.}~\bibnamefont
  {Hemmerling}}, \bibinfo {author} {\bibfnamefont {E.}~\bibnamefont {Chae}},
  \bibinfo {author} {\bibfnamefont {L.}~\bibnamefont {Anderegg}}, \bibinfo
  {author} {\bibfnamefont {G.}~\bibnamefont {Drayna}}, \bibinfo {author}
  {\bibfnamefont {N.}~\bibnamefont {Hutzler}}, \bibinfo {author} {\bibfnamefont
  {A.}~\bibnamefont {Collopy}}, \bibinfo {author} {\bibfnamefont
  {J.}~\bibnamefont {Ye}}, \bibinfo {author} {\bibfnamefont {W.}~\bibnamefont
  {Ketterle}},\ and\ \bibinfo {author} {\bibfnamefont {J.}~\bibnamefont
  {Doyle}},\ }\bibfield  {title} {\bibinfo {title} {Laser slowing of caf
  molecules to near the capture velocity of a molecular mot},\ }\href
  {https://doi.org/10.1088/0953-4075/49/17/174001} {\bibfield  {journal}
  {\bibinfo  {journal} {Journal of Physics B: Atomic, Molecular and Optical
  Physics}\ }\textbf {\bibinfo {volume} {49}} (\bibinfo {year}
  {2016})}\BibitemShut {NoStop}%
\bibitem [{\citenamefont {{V{\'a}zquez-Carson}}\ \emph
  {et~al.}(2022)\citenamefont {{V{\'a}zquez-Carson}}, \citenamefont {{Sun}},
  \citenamefont {{Dai}}, \citenamefont {{Mitra}},\ and\ \citenamefont
  {{Zelevinsky}}}]{Zelevinsky2022}%
  \BibitemOpen
  \bibfield  {author} {\bibinfo {author} {\bibfnamefont {S.~F.}\ \bibnamefont
  {{V{\'a}zquez-Carson}}}, \bibinfo {author} {\bibfnamefont {Q.}~\bibnamefont
  {{Sun}}}, \bibinfo {author} {\bibfnamefont {J.}~\bibnamefont {{Dai}}},
  \bibinfo {author} {\bibfnamefont {D.}~\bibnamefont {{Mitra}}},\ and\ \bibinfo
  {author} {\bibfnamefont {T.}~\bibnamefont {{Zelevinsky}}},\ }\bibfield
  {title} {\bibinfo {title} {{Direct laser cooling of calcium monohydride
  molecules}},\ }\href {https://doi.org/10.1088/1367-2630/ac806c} {\bibfield
  {journal} {\bibinfo  {journal} {New Journal of Physics}\ }\textbf {\bibinfo
  {volume} {24}},\ \bibinfo {eid} {083006} (\bibinfo {year} {2022})},\ \Eprint
  {https://arxiv.org/abs/2203.04841} {arXiv:2203.04841 [physics.atom-ph]}
  \BibitemShut {NoStop}%
\bibitem [{\citenamefont {Langen}\ \emph {et~al.}(2024)\citenamefont {Langen},
  \citenamefont {Valtolina}, \citenamefont {Wang},\ and\ \citenamefont
  {Ye}}]{Tim2024}%
  \BibitemOpen
  \bibfield  {author} {\bibinfo {author} {\bibfnamefont {T.}~\bibnamefont
  {Langen}}, \bibinfo {author} {\bibfnamefont {G.}~\bibnamefont {Valtolina}},
  \bibinfo {author} {\bibfnamefont {D.}~\bibnamefont {Wang}},\ and\ \bibinfo
  {author} {\bibfnamefont {J.}~\bibnamefont {Ye}},\ }\bibfield  {title}
  {\bibinfo {title} {Quantum state manipulation and cooling of ultracold
  molecules},\ }\href {https://doi.org/10.1038/s41567-024-02423-1} {\bibfield
  {journal} {\bibinfo  {journal} {Nature Physics}\ }\textbf {\bibinfo {volume}
  {20}},\ \bibinfo {pages} {702} (\bibinfo {year} {2024})}\BibitemShut
  {NoStop}%
\bibitem [{\citenamefont {Wright}\ \emph {et~al.}(2022)\citenamefont {Wright},
  \citenamefont {Doppelbauer}, \citenamefont {Hofs{\"a}ss}, \citenamefont
  {Schewe}, \citenamefont {Sartakov}, \citenamefont {Meijer}, \citenamefont
  {Truppe},\ and\ \citenamefont {Haber}}]{Wright2022}%
  \BibitemOpen
  \bibfield  {author} {\bibinfo {author} {\bibfnamefont {S.}~\bibnamefont
  {Wright}}, \bibinfo {author} {\bibfnamefont {M.}~\bibnamefont {Doppelbauer}},
  \bibinfo {author} {\bibfnamefont {S.}~\bibnamefont {Hofs{\"a}ss}}, \bibinfo
  {author} {\bibfnamefont {H.~C.}\ \bibnamefont {Schewe}}, \bibinfo {author}
  {\bibfnamefont {B.~G.}\ \bibnamefont {Sartakov}}, \bibinfo {author}
  {\bibfnamefont {G.}~\bibnamefont {Meijer}}, \bibinfo {author} {\bibfnamefont
  {S.}~\bibnamefont {Truppe}},\ and\ \bibinfo {author} {\bibfnamefont
  {F.}~\bibnamefont {Haber}},\ }\bibfield  {title} {\bibinfo {title} {Cryogenic
  buffer gas beams of alf, caf, mgf, ybf, al, ca, yb and no – a comparison},\
  }\href {https://api.semanticscholar.org/CorpusID:252222406} {\bibfield
  {journal} {\bibinfo  {journal} {Molecular Physics}\ }\textbf {\bibinfo
  {volume} {121}} (\bibinfo {year} {2022})}\BibitemShut {NoStop}%
\bibitem [{\citenamefont {Truppe}\ \emph
  {et~al.}(2018{\natexlab{b}})\citenamefont {Truppe}, \citenamefont {Hambach},
  \citenamefont {Skoff}, \citenamefont {Bulleid}, \citenamefont {Bumby},
  \citenamefont {Hendricks}, \citenamefont {Hinds}, \citenamefont {E.},\ and\
  \citenamefont {Tarbutt}}]{Truppe2018}%
  \BibitemOpen
  \bibfield  {author} {\bibinfo {author} {\bibfnamefont {S.}~\bibnamefont
  {Truppe}}, \bibinfo {author} {\bibfnamefont {M.}~\bibnamefont {Hambach}},
  \bibinfo {author} {\bibfnamefont {S.~M.}\ \bibnamefont {Skoff}}, \bibinfo
  {author} {\bibfnamefont {N.}~\bibnamefont {Bulleid}}, \bibinfo {author}
  {\bibfnamefont {J.~S.}\ \bibnamefont {Bumby}}, \bibinfo {author}
  {\bibfnamefont {R.~J.}\ \bibnamefont {Hendricks}}, \bibinfo {author}
  {\bibfnamefont {E.~A.}\ \bibnamefont {Hinds}}, \bibinfo {author}
  {\bibfnamefont {S.~B.}\ \bibnamefont {E.}},\ and\ \bibinfo {author}
  {\bibfnamefont {M.~R.}\ \bibnamefont {Tarbutt}},\ }\bibfield  {title}
  {\bibinfo {title} {A buffer gas beam source for short, intense and slow
  molecular pulses},\ }\href {https://doi.org/10.1080/09500340.2017.1384516}
  {\bibfield  {journal} {\bibinfo  {journal} {Journal of Modern Optics}\
  }\textbf {\bibinfo {volume} {65}},\ \bibinfo {pages} {648} (\bibinfo {year}
  {2018}{\natexlab{b}})},\ \Eprint
  {https://arxiv.org/abs/https://doi.org/10.1080/09500340.2017.1384516}
  {https://doi.org/10.1080/09500340.2017.1384516} \BibitemShut {NoStop}%
\bibitem [{\citenamefont {Ganesan-Santhi}\ \emph {et~al.}(2024)\citenamefont
  {Ganesan-Santhi}, \citenamefont {Frye}, \citenamefont {Gronowski},\ and\
  \citenamefont {Tomza}}]{Sangami2024}%
  \BibitemOpen
  \bibfield  {author} {\bibinfo {author} {\bibfnamefont {S.}~\bibnamefont
  {Ganesan-Santhi}}, \bibinfo {author} {\bibfnamefont {M.~D.}\ \bibnamefont
  {Frye}}, \bibinfo {author} {\bibfnamefont {M.}~\bibnamefont {Gronowski}},\
  and\ \bibinfo {author} {\bibfnamefont {M.}~\bibnamefont {Tomza}},\ }\bibfield
   {title} {\bibinfo {title} {Interactions and cold collisions of alf in the
  ground and excited electronic states with he},\ }\href
  {https://arxiv.org/abs/2405.03276} {\bibfield  {journal} {\bibinfo  {journal}
  {Arxiv}\ } (\bibinfo {year} {2024})},\ \Eprint
  {https://arxiv.org/abs/2405.03276} {arXiv:2405.03276 [physics.chem-ph]}
  \BibitemShut {NoStop}%
\bibitem [{\citenamefont {Karra}\ \emph {et~al.}(2022)\citenamefont {Karra},
  \citenamefont {Cretu}, \citenamefont {Friedrich}, \citenamefont {Truppe},
  \citenamefont {Meijer},\ and\ \citenamefont {P\'erez-R\'{\i}os}}]{Karra2022}%
  \BibitemOpen
  \bibfield  {author} {\bibinfo {author} {\bibfnamefont {M.}~\bibnamefont
  {Karra}}, \bibinfo {author} {\bibfnamefont {M.~T.}\ \bibnamefont {Cretu}},
  \bibinfo {author} {\bibfnamefont {B.}~\bibnamefont {Friedrich}}, \bibinfo
  {author} {\bibfnamefont {S.}~\bibnamefont {Truppe}}, \bibinfo {author}
  {\bibfnamefont {G.}~\bibnamefont {Meijer}},\ and\ \bibinfo {author}
  {\bibfnamefont {J.}~\bibnamefont {P\'erez-R\'{\i}os}},\ }\bibfield  {title}
  {\bibinfo {title} {Dynamics of translational and rotational thermalization of
  alf molecules via collisions with cryogenic helium},\ }\href
  {https://doi.org/10.1103/PhysRevA.105.022808} {\bibfield  {journal} {\bibinfo
   {journal} {Phys. Rev. A}\ }\textbf {\bibinfo {volume} {105}},\ \bibinfo
  {pages} {022808} (\bibinfo {year} {2022})}\BibitemShut {NoStop}%
\bibitem [{\citenamefont {Skoff}\ \emph {et~al.}(2011)\citenamefont {Skoff},
  \citenamefont {Hendricks}, \citenamefont {Sinclair}, \citenamefont {Hudson},
  \citenamefont {Segal}, \citenamefont {Sauer}, \citenamefont {Hinds},\ and\
  \citenamefont {Tarbutt}}]{Skoff2011}%
  \BibitemOpen
  \bibfield  {author} {\bibinfo {author} {\bibfnamefont {S.~M.}\ \bibnamefont
  {Skoff}}, \bibinfo {author} {\bibfnamefont {R.~J.}\ \bibnamefont
  {Hendricks}}, \bibinfo {author} {\bibfnamefont {C.~D.~J.}\ \bibnamefont
  {Sinclair}}, \bibinfo {author} {\bibfnamefont {J.~J.}\ \bibnamefont
  {Hudson}}, \bibinfo {author} {\bibfnamefont {D.~M.}\ \bibnamefont {Segal}},
  \bibinfo {author} {\bibfnamefont {B.~E.}\ \bibnamefont {Sauer}}, \bibinfo
  {author} {\bibfnamefont {E.~A.}\ \bibnamefont {Hinds}},\ and\ \bibinfo
  {author} {\bibfnamefont {M.~R.}\ \bibnamefont {Tarbutt}},\ }\bibfield
  {title} {\bibinfo {title} {Diffusion, thermalization, and optical pumping of
  ybf molecules in a cold buffer-gas cell},\ }\href
  {https://doi.org/10.1103/PhysRevA.83.023418} {\bibfield  {journal} {\bibinfo
  {journal} {Phys. Rev. A}\ }\textbf {\bibinfo {volume} {83}},\ \bibinfo
  {pages} {023418} (\bibinfo {year} {2011})}\BibitemShut {NoStop}%
\bibitem [{\citenamefont {Wang}\ \emph {et~al.}(2023)\citenamefont {Wang},
  \citenamefont {Julian}, \citenamefont {Ibrahim}, \citenamefont {Chin},
  \citenamefont {Bhattiprolu}, \citenamefont {Franco},\ and\ \citenamefont
  {P{\'e}rez-R{\'\i}os}}]{database}%
  \BibitemOpen
  \bibfield  {author} {\bibinfo {author} {\bibfnamefont {Y.}~\bibnamefont
  {Wang}}, \bibinfo {author} {\bibfnamefont {D.}~\bibnamefont {Julian}},
  \bibinfo {author} {\bibfnamefont {M.~A.}\ \bibnamefont {Ibrahim}}, \bibinfo
  {author} {\bibfnamefont {C.}~\bibnamefont {Chin}}, \bibinfo {author}
  {\bibfnamefont {S.}~\bibnamefont {Bhattiprolu}}, \bibinfo {author}
  {\bibfnamefont {E.}~\bibnamefont {Franco}},\ and\ \bibinfo {author}
  {\bibfnamefont {J.}~\bibnamefont {P{\'e}rez-R{\'\i}os}},\ }\bibfield  {title}
  {\bibinfo {title} {The database of spectroscopic constants of diatomic
  molecules (dscdm): A dynamic and user-friendly interface for molecular
  physics and spectroscopy},\ }\href
  {https://doi.org/https://doi.org/10.1016/j.jms.2023.111848} {\bibfield
  {journal} {\bibinfo  {journal} {Journal of Molecular Spectroscopy}\ }\textbf
  {\bibinfo {volume} {398}},\ \bibinfo {pages} {111848} (\bibinfo {year}
  {2023})}\BibitemShut {NoStop}%
\bibitem [{\citenamefont {Schuchardt}\ \emph {et~al.}(2007)\citenamefont
  {Schuchardt}, \citenamefont {Didier}, \citenamefont {Elsethagen},
  \citenamefont {Sun}, \citenamefont {Gurumoorthi}, \citenamefont {Chase},
  \citenamefont {Li},\ and\ \citenamefont {Windus}}]{basisset}%
  \BibitemOpen
  \bibfield  {author} {\bibinfo {author} {\bibfnamefont {K.~L.}\ \bibnamefont
  {Schuchardt}}, \bibinfo {author} {\bibfnamefont {B.~T.}\ \bibnamefont
  {Didier}}, \bibinfo {author} {\bibfnamefont {T.}~\bibnamefont {Elsethagen}},
  \bibinfo {author} {\bibfnamefont {L.}~\bibnamefont {Sun}}, \bibinfo {author}
  {\bibfnamefont {V.}~\bibnamefont {Gurumoorthi}}, \bibinfo {author}
  {\bibfnamefont {J.}~\bibnamefont {Chase}}, \bibinfo {author} {\bibfnamefont
  {J.}~\bibnamefont {Li}},\ and\ \bibinfo {author} {\bibfnamefont {T.~L.}\
  \bibnamefont {Windus}},\ }\bibfield  {title} {\bibinfo {title} {Basis set
  exchange: A community database for computational sciences},\ }\href
  {https://doi.org/10.1021/ci600510j} {\bibfield  {journal} {\bibinfo
  {journal} {Journal of Chemical Information and Modeling}\ }\textbf {\bibinfo
  {volume} {47}},\ \bibinfo {pages} {1045} (\bibinfo {year}
  {2007})}\BibitemShut {NoStop}%
\bibitem [{\citenamefont {Denis-Alpizar}\ \emph {et~al.}(2024)\citenamefont
  {Denis-Alpizar}, \citenamefont {Zanchet},\ and\ \citenamefont
  {Stoecklin}}]{Alpizar2024}%
  \BibitemOpen
  \bibfield  {author} {\bibinfo {author} {\bibfnamefont {O.}~\bibnamefont
  {Denis-Alpizar}}, \bibinfo {author} {\bibfnamefont {A.}~\bibnamefont
  {Zanchet}},\ and\ \bibinfo {author} {\bibfnamefont {T.}~\bibnamefont
  {Stoecklin}},\ }\bibfield  {title} {\bibinfo {title} {Quantum study of the
  rovibrational relaxation of hf by collision with 4he on a new potential
  energy surface},\ }\href {https://doi.org/10.1039/D3CP05606F} {\bibfield
  {journal} {\bibinfo  {journal} {Phys. Chem. Chem. Phys.}\ }\textbf {\bibinfo
  {volume} {26}},\ \bibinfo {pages} {13432} (\bibinfo {year}
  {2024})}\BibitemShut {NoStop}%
\bibitem [{\citenamefont {Bop}\ \emph {et~al.}(2019)\citenamefont {Bop},
  \citenamefont {Faye},\ and\ \citenamefont {Hammami}}]{Bop2019}%
  \BibitemOpen
  \bibfield  {author} {\bibinfo {author} {\bibfnamefont {C.~T.}\ \bibnamefont
  {Bop}}, \bibinfo {author} {\bibfnamefont {N.}~\bibnamefont {Faye}},\ and\
  \bibinfo {author} {\bibfnamefont {K.}~\bibnamefont {Hammami}},\ }\bibfield
  {title} {\bibinfo {title} {Sodium hydride nah(x1s+) in collision with helium
  he(1s) at low temperature: Potential energy surface and rotational rate
  coefficients},\ }\href
  {https://doi.org/https://doi.org/10.1016/j.chemphys.2018.11.021} {\bibfield
  {journal} {\bibinfo  {journal} {Chemical Physics}\ }\textbf {\bibinfo
  {volume} {519}},\ \bibinfo {pages} {21} (\bibinfo {year} {2019})}\BibitemShut
  {NoStop}%
\bibitem [{\citenamefont {Arthurs}\ \emph {et~al.}(1960)\citenamefont
  {Arthurs}, \citenamefont {Dalgarno},\ and\ \citenamefont
  {Bates}}]{Arthurs1960}%
  \BibitemOpen
  \bibfield  {author} {\bibinfo {author} {\bibfnamefont {A.~M.}\ \bibnamefont
  {Arthurs}}, \bibinfo {author} {\bibfnamefont {A.}~\bibnamefont {Dalgarno}},\
  and\ \bibinfo {author} {\bibfnamefont {D.~R.}\ \bibnamefont {Bates}},\
  }\bibfield  {title} {\bibinfo {title} {The theory of scattering by a rigid
  rotator},\ }\href {https://doi.org/10.1098/rspa.1960.0125} {\bibfield
  {journal} {\bibinfo  {journal} {Proceedings of the Royal Society of London.
  Series A. Mathematical and Physical Sciences}\ }\textbf {\bibinfo {volume}
  {256}},\ \bibinfo {pages} {540} (\bibinfo {year} {1960})},\ \Eprint
  {https://arxiv.org/abs/https://royalsocietypublishing.org/doi/pdf/10.1098/rspa.1960.0125}
  {https://royalsocietypublishing.org/doi/pdf/10.1098/rspa.1960.0125}
  \BibitemShut {NoStop}%
\bibitem [{\citenamefont {Mott}\ and\ \citenamefont {Massey}(1933)}]{mottbook}%
  \BibitemOpen
  \bibfield  {author} {\bibinfo {author} {\bibfnamefont {N.~F. N.~F.}\
  \bibnamefont {Mott}}\ and\ \bibinfo {author} {\bibfnamefont {H.}~\bibnamefont
  {Massey}},\ }\href@noop {} {\emph {\bibinfo {title} {The theory of atomic
  collisions}}},\ The International series of monographs on physics\ (\bibinfo
  {publisher} {Clarendon Press},\ \bibinfo {address} {Oxford},\ \bibinfo {year}
  {1933})\BibitemShut {NoStop}%
\bibitem [{\citenamefont {Messiah}(1961)}]{messiah61}%
  \BibitemOpen
  \bibfield  {author} {\bibinfo {author} {\bibfnamefont {A.}~\bibnamefont
  {Messiah}},\ }\href@noop {} {\emph {\bibinfo {title} {Quantum Mechanics
  Volume II}}}\ (\bibinfo  {publisher} {Elsevier Science B.V.},\ \bibinfo
  {year} {1961})\BibitemShut {NoStop}%
\bibitem [{\citenamefont {Hutson}\ and\ \citenamefont {{Le
  Sueur}}(2019)}]{HUTSON2019}%
  \BibitemOpen
  \bibfield  {author} {\bibinfo {author} {\bibfnamefont {J.~M.}\ \bibnamefont
  {Hutson}}\ and\ \bibinfo {author} {\bibfnamefont {C.~R.}\ \bibnamefont {{Le
  Sueur}}},\ }\bibfield  {title} {\bibinfo {title} {molscat: A program for
  non-reactive quantum scattering calculations on atomic and molecular
  collisions},\ }\href
  {https://doi.org/https://doi.org/10.1016/j.cpc.2019.02.014} {\bibfield
  {journal} {\bibinfo  {journal} {Computer Physics Communications}\ }\textbf
  {\bibinfo {volume} {241}},\ \bibinfo {pages} {9} (\bibinfo {year}
  {2019})}\BibitemShut {NoStop}%
\bibitem [{\citenamefont {Manolopoulos}(1986)}]{Manolopoulos1986}%
  \BibitemOpen
  \bibfield  {author} {\bibinfo {author} {\bibfnamefont {D.~E.}\ \bibnamefont
  {Manolopoulos}},\ }\bibfield  {title} {\bibinfo {title} {An improved log
  derivative method for inelastic scattering},\ }\href
  {https://doi.org/10.1063/1.451472} {\bibfield  {journal} {\bibinfo  {journal}
  {The Journal of Chemical Physics}\ }\textbf {\bibinfo {volume} {85}},\
  \bibinfo {pages} {6425} (\bibinfo {year} {1986})},\ \Eprint
  {https://arxiv.org/abs/https://pubs.aip.org/aip/jcp/article-pdf/85/11/6425/18962304/6425\_1\_online.pdf}
  {https://pubs.aip.org/aip/jcp/article-pdf/85/11/6425/18962304/6425\_1\_online.pdf}
  \BibitemShut {NoStop}%
\bibitem [{\citenamefont {Alexander}(1984)}]{Alexander1984}%
  \BibitemOpen
  \bibfield  {author} {\bibinfo {author} {\bibfnamefont {M.~H.}\ \bibnamefont
  {Alexander}},\ }\bibfield  {title} {\bibinfo {title} {Hybrid quantum
  scattering algorithms for long‐range potentials},\ }\href
  {https://doi.org/10.1063/1.447420} {\bibfield  {journal} {\bibinfo  {journal}
  {The Journal of Chemical Physics}\ }\textbf {\bibinfo {volume} {81}},\
  \bibinfo {pages} {4510} (\bibinfo {year} {1984})},\ \Eprint
  {https://arxiv.org/abs/https://pubs.aip.org/aip/jcp/article-pdf/81/10/4510/18951032/4510\_1\_online.pdf}
  {https://pubs.aip.org/aip/jcp/article-pdf/81/10/4510/18951032/4510\_1\_online.pdf}
  \BibitemShut {NoStop}%
\bibitem [{\citenamefont {Montero}\ and\ \citenamefont
  {P{\'e}rez-R{\'\i}os}(2014)}]{Montero}%
  \BibitemOpen
  \bibfield  {author} {\bibinfo {author} {\bibfnamefont {S.}~\bibnamefont
  {Montero}}\ and\ \bibinfo {author} {\bibfnamefont {J.}~\bibnamefont
  {P{\'e}rez-R{\'\i}os}},\ }\bibfield  {title} {\bibinfo {title} {Rotational
  relaxation in molecular hydrogen and deuterium: Theory versus acoustic
  experiments},\ }\href {https://doi.org/10.1063/1.4895398} {\bibfield
  {journal} {\bibinfo  {journal} {The Journal of Chemical Physics}\ }\textbf
  {\bibinfo {volume} {141}},\ \bibinfo {pages} {114301} (\bibinfo {year}
  {2014})},\ \Eprint
  {https://arxiv.org/abs/https://pubs.aip.org/aip/jcp/article-pdf/doi/10.1063/1.4895398/15483890/114301\_1\_online.pdf}
  {https://pubs.aip.org/aip/jcp/article-pdf/doi/10.1063/1.4895398/15483890/114301\_1\_online.pdf}
  \BibitemShut {NoStop}%
\end{thebibliography}%


\end{document}